\documentclass[twocolumn,amsmath,amssymb,aps,prd,10pt,superscriptaddress,nofootinbib,noeprint,preprintnumbers,floatfix,notitlepage]{revtex4-2}

\newcommand{\SLJ}[3]{\ensuremath{{\:\!}^{{#1}\!}{#2}_{#3}}}
\newcommand{\SLJc}[3]{\ensuremath{  \{  {\:\!}^{{#1}\!}{#2}_{#3}  \}  }}

\usepackage{amsmath,amssymb,amsfonts,braket,slashed,mathrsfs,dsfont}
\usepackage{bm}
\usepackage{slashed}
\usepackage{bbold}
\usepackage{tensor}
\usepackage{diagbox}

\usepackage{lipsum}

\newcommand{\eg}{\textit{e.g.}~}
\newcommand{\ie}{\textit{i.e.}~}

\newcommand{\CM}{c.m. }
\newcommand{\thr}{{\sf thr}}
\newcommand{\mev}{~\mathrm{MeV}}

\usepackage{amsmath,amssymb,amsfonts,braket,slashed,mathrsfs,dsfont}

\newcommand{\Ecm}{\ensuremath{E_\mathrm{cm}}}
\newcommand{\chisqadd}{\ensuremath{\chi^2_\mathrm{add}}}

\DeclareMathOperator{\Kmatb}{\mathbf{K}}
\DeclareMathOperator{\Kmat}{K}

\DeclareMathOperator{\Real}{Re}
\DeclareMathOperator{\Imag}{Im}

\usepackage{lineno}

\pdfoutput=1

\usepackage{graphicx}
\usepackage{dcolumn}
\usepackage{bm}
\usepackage{mathtools}
\usepackage{multirow}
\usepackage{amssymb}
\usepackage{amsmath}
\usepackage{amsfonts}
\usepackage{xcolor}
\usepackage{tabularx}
\usepackage{array}
\usepackage[utf8]{inputenc}
\usepackage{hyperref}
\usepackage[letterspace=-10]{microtype}
\usepackage{soul}

\definecolor{jlab_red}{RGB}{192,39,45}
\definecolor{jlab_orange}{RGB}{249,102,0}
\definecolor{jlab_blue}{RGB}{47,122,121}
\definecolor{jlab_green}{RGB}{65,125,10}

\hypersetup{%
pdftitle = {Coupled-channel scattering of $DD, DD^*$ and $D^* D^*$ in isospin-$1$ from lattice QCD},
pdfsubject = {QCD},
pdfkeywords = {QCD, Hadron, Charm, Lattice, Meson, Scattering},
pdfauthor = {Hadron Spectrum Collaboration},
colorlinks = {true},
filecolor = {black},
linkcolor = {jlab_blue},
menucolor = {black},
citecolor = {jlab_green},
urlcolor = {jlab_green},
}{}

\begin{document}

\title{
 Coupled-channel scattering of $DD, DD^*$ and $D^*D^*$ in isospin-$1$ from lattice QCD
}

\author{Nelson Pitanga Lachini}
 \email{np612@cam.ac.uk}
\affiliation{
Department of Applied Mathematics and Theoretical Physics, Center for Mathematical Sciences,\\University of Cambridge, Wilberforce Road, Cambridge, CB3 0WA, United Kingdom
}
\author{Christopher E. Thomas}
\email{c.e.thomas@damtp.cam.ac.uk}
\affiliation{
Department of Applied Mathematics and Theoretical Physics, Center for Mathematical Sciences,\\University of Cambridge, Wilberforce Road, Cambridge, CB3 0WA, United Kingdom
}
\author{David J. Wilson}
\email{d.j.wilson@damtp.cam.ac.uk}
\affiliation{
Department of Applied Mathematics and Theoretical Physics, Center for Mathematical Sciences,\\University of Cambridge, Wilberforce Road, Cambridge, CB3 0WA, United Kingdom
}
\collaboration{for the Hadron Spectrum Collaboration}

\begin{abstract}
    \noindent
    The first coupled-channel determination of two-body $D$ and $D^*$ scattering amplitudes in isospin-$1$ from lattice quantum chromodynamics is presented.
    Using three lattice volumes at $m_\pi \approx 391\mev$, finite-volume energies relevant for the channels of interest are determined.
    Through the Lüscher formalism, these energies are used to constrain amplitudes of coupled $J^P=0^+$ $DD - D^*D^*$, $J^P=1^+$ $DD^* - D^*D^*$ and $J^P=2^+$  $DD - DD^* - D^*D^*$ scattering.
    All channels feature weakly repulsive interactions in $S$ wave, except for a weak $D^*D^*$ attraction  in $J^P=0^+$.
    No amplitude singularities corresponding to physical states are found.
    Some of these amplitudes will be a necessary component of future lattice QCD analyses of $DD\pi$ and $DD^*$ scattering taking into account three-body and left-hand cut effects.
\end{abstract}

\maketitle

\section{ \label{sec:intro}Introduction}

In recent years much effort has been put into investigating the landscape of exotic hadrons.
This is evidenced by the sheer number of such states detected at the LHCb, BESIII, Belle, BaBar and other experiments~\cite{ParticleDataGroup:2024cfk, Johnson:2024omq}.
In particular, the observation of the exotic $T_{cc}^+(3875)$ in $D^0D^0\pi^+$ scattering has drawn considerable attention towards the doubly charmed sector~\cite{LHCb:2021vvq,LHCb:2021auc}. It arises as a narrow resonance near the $DD^*$ threshold, which the experimental data strongly constrains to be a ${I(J^P)=0(1^+)}$ state, where $I,J$ and $P$ are the isospin, angular momentum and parity quantum numbers, respectively.

The theoretical predictions and postdictions of doubly charmed exotics have been largely driven by extensions of the quark model, heavy quark spin symmetry, and effective theory methods, see for example Refs.~\cite{Chen:2022asf, Neubert:1993mb, Meng:2022ozq}.
Among these, a state with the features of the observed $T_{cc}^+$ is generally present but a description of its potential symmetry partners can vary depending on the approach.
For example, within a particular framework, either a $DD^*$, ${I(J^P)=0(1^+)}$ or a ${1(1^+)}$ state can arise, and if so, it is accompanied by a $D^*D^*$ heavy spin-symmetry partner in ${I(J^P)=0(1^+)}$ or ${1(2^+)}$, respectively~\cite{Albaladejo:2021vln}.
In another instance, an ${I=1}$ state is predicted in coupled-channel $DD, D^*D^*$, $J^P=0^+$~\cite{Ortega:2022efc}.
Conversely, examples exist where no state beyond the $T_{cc}^+$ can be found across $I=0,1$ and $J^P=0^+,1^+,2^+$~\cite{Yang:2019itm}.
Such theoretical approaches are useful to guide experimental searches and provide an understanding of the physical mechanisms behind the strong interaction.
Nevertheless, it is also important to extract information directly from the theory underlying the strong interactions, quantum chromodynamics (QCD).

Lattice QCD provides quantitative predictions based on systematically improvable calculations of QCD, fully taking into account its strongly-coupled dynamics.
This approach is based on the discretization of spacetime into a hypercubic lattice, which allows for the Monte Carlo sampling of the path integral and thus the nonperturbative computation of correlation functions.
The spectrum in a finite and periodic spatial volume is discrete, and the energy eigenstates can be extracted from correlation functions.
Finally, the relation between the infinite-volume scattering amplitudes and the finite-volume spectrum provides a quantitative description of two-hadron systems~\cite{Luscher:1990ux,Rummukainen:1995vs,Bedaque:2004kc,Kim:2005gf,He:2005ey,Lage:2009zv,Fu:2011xz,Leskovec:2012gb,Gockeler:2012yj,Hansen:2012tf,Briceno:2012yi,Guo:2012hv,Briceno:2014oea}.

Due to the increasing interest in the $T_{cc}^+$ state, several lattice investigations of the $J^P=1^+$, $DD^*$ channel have been recently performed for $I=0$~\cite{Junnarkar:2018twb, Padmanath:2022cvl, Lyu:2023xro, Collins:2024sfi}, including a calculation in the ${DD^*,D^*D^*}$ coupled-channel region~\cite{Whyte:2024ihh}.
Some lattice studies exist for the case of $I=1$, but where focus was given to elastic $J^P=1^+$, $DD^*$ scattering~\cite{Chen:2022vpo,Meng:2024kkp}.
The elastic $J^P=0^+$, $DD$ channel was studied in Ref.~\cite{Shi:2025ogt}, and also using the HALQCD method in Ref.~\cite{Ikeda:2013vwa}.
Overall the interactions in $I=1$ were observed to be small and repulsive, with no signal of potential states.
As of now, there are no lattice calculations regarding $D^*D^*, J^P=2^+$ scattering in $I=1$.
Furthermore, there are also no works that take into account coupled-channel effects in this $I=1$ sector for any $J^P$.

In this article, we present the first coupled-channel analysis of the two-body channels involving $D$ and $D^*$ with overall $J^P=0^+,1^+,2^+$ in $I=1$.
More precisely, scattering amplitudes for ${DD,D^*D^*}$ in ${J^P=0^+}$, ${DD^*,D^*D^*}$ in ${J^P=1^+}$, and ${DD,DD^*,D^*D^*}$ in ${J^P=2^+}$ are extracted.
The interactions are observed to be overall weakly-repulsive across all channels, and no states are found.
Besides extending the picture of QCD calculations of the doubly charmed sector, we note that extracting the $DD$ amplitude in $I=1$ is a necessary ingredient for a future three-body analysis of $DD\pi$ scattering and for treating the $DD^*$ left-hand cut in $I=0$~\cite{Hansen:2024ffk,Dawid:2024dgy}.
In these analyses, both two-body $DD$, $I=1$ and $D\pi$, $I=1/2$ amplitudes are used to extract the three-body $DD\pi$ amplitude in $I=0$.
In the case where $D^*$ is stable, the three-body approach can also be used to treat the left-hand cut appearing in two-body $DD^*$, $I=0$ scattering with the $I=1$ and $I=1/2$ amplitudes as inputs.

We use the same setup and methods as in Ref.~\cite{Whyte:2024ihh}, where a virtual bound-state corresponding to the $T_{cc}^+$ was found near the $DD^*$ threshold in $I=0$, ${J^P=1^+}$, together with a resonance pole just below the $D^*D^*$ threshold.
In particular, we use the same lattices, where the quark masses are tuned such that $m_\pi \approx 391\mev$, and so the $D^*$ is stable.
In this way, $DD^*$ and $D^*D^*$ can be treated as two-body problems below the $DD\pi$ threshold.
We also note the presence of left-hand cuts near the lowest thresholds occurring here, analogous to those in the $I=0$ channel~\cite{Whyte:2024ihh, Collins:2024sfi, Meng:2024kkp}.
We do not account for the potential effects of such cuts here.
Proposals on how to explicitly deal with those effects have recently appeared in the literature~\cite{Raposo:2023oru,Raposo:2025dkb,Meng:2023bmz,Hansen:2024ffk,Dawid:2024oey,Bubna:2024izx}, and analyses or reanalyses of lattice data have been performed for the doubly-charmed sector in $I=0$~\cite{Collins:2024sfi,Dawid:2024dgy,Gil-Dominguez:2024zmr,Prelovsek:2025vbr} and in $I=1$~\cite{Meng:2024kkp}.

The rest of this manuscript is organized as follows.
In Sec.~\ref{sec:details} we provide an overview of the allowed $D^{(*)}D^{(*)}$ partial-waves in $I=1$ and briefly review the setup used.
In Sec.~\ref{sec:fvspectra}, we present the finite-volume spectra extracted from the lattice and comment on their contribution to the amplitudes of interest.
The scattering analysis results are presented in Sec.~\ref{sec:scattering}.
We interpret the results in Sec.~\ref{sec:interpret} and conclude in Sec.~\ref{sec:conclusions}.   

\section{\label{sec:details} Calculation Details}

\subsection{Partial waves}

We first consider the relevant infinite-volume two-hadron states with $I=1$ and positive parity ($P=+$).
The $DD$ and $D^*D^*$ states are constrained to be overall symmetric due to Bose symmetry.
As the $I=1$ isospin wavefunction is symmetric, the product of the spin and spatial ones for $DD$ and $D^*D^*$ is restricted to be symmetric.
While for $DD$ the spin $S=0$ wave is symmetric, $D^*D^*$ can be decomposed into an antisymmetric $S=1$ and symmetric $S=0,2$ contributions.
Note that, the allowed $DD^*$ states are only restricted by parity as the hadrons are non-identical.
In Table~\ref{tab:partialwavebyJp}, the allowed partial waves are labelled using the spectroscopic notation.
The leading negative-parity partial waves are shown in Appendix~\ref{apx:negparity}.

\subsection{Lattice setup and methods}

\begin{figure*}[!t]
    \centering
    \includegraphics[width=0.44\textwidth, trim={2.5mm 2mm 2.5mm 2mm},clip]{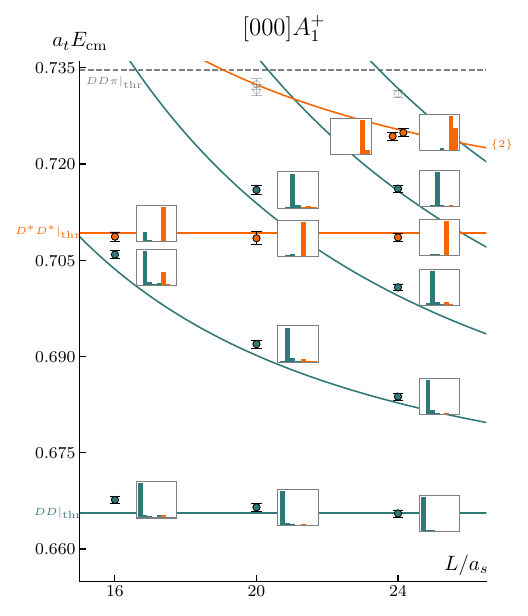}
    \includegraphics[width=0.415\textwidth, trim={2.5mm 2mm 2.5mm 2mm},clip]{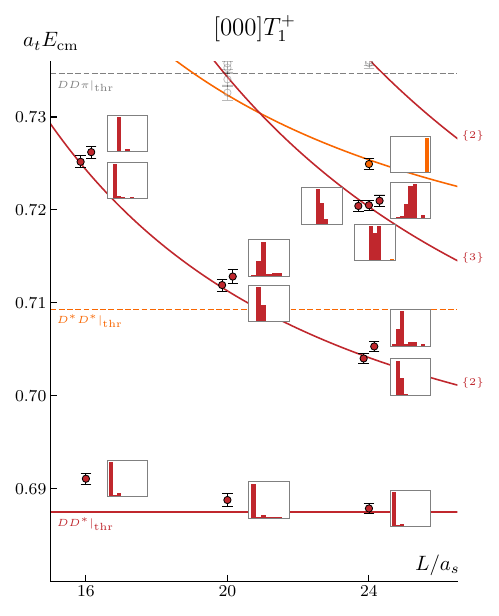}
    \includegraphics[width=0.0985\textwidth,  trim={2.5mm 2mm 0mm 2mm},clip]{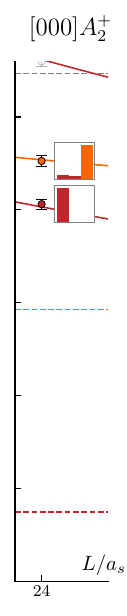}

    \caption{\label{fig:fvspectra1}
    Finite-volume spectrum for rest-frame irreps $[000]A_1^+$, $[000]T_1^+$ and $[000]A_2^+$.
    The points are the computed lattice energies and error bars are statistical with the additional systematic uncertainty added in quadrature, as described in the text, and colored based on the magnitude of their largest operator-state overlap.
    The translucent data points were not used in the following scattering analysis (see also Appendix~\ref{apx:operatortables}).
    The inset histograms represent the magnitudes of the normalized operator-state overlaps associated with the energy level they are next to, as described in the text.
    The noninteracting $DD$ (blue), $DD^*$ (orange) and $D^*D^*$ (red) finite-volume energies are represented by solid lines.
    When the degeneracy of a noninteracting level is larger than one, this is indicated by a number in curly brackets, and the corresponding lattice levels are displaced horizontally to facilitate visualization.
    Following the same color scheme, the opening of kinematic thresholds is represented by horizontal dashed lines, where the $D_0^*D$ threshold ($a_t \Ecm \approx 0.734$) has been omitted due to its proximity to the $DD\pi$ threshold (see Table~\ref{tab:hadmasses}).
    }
\end{figure*}

\begin{table}[!t]
    \centering
    \setlength{\tabcolsep}{0.5em}{\renewcommand{\arraystretch}{1.5}{
    \begin{tabular}{c||c|c|c}
    $J^P$         & $DD$      & $DD^*$                   & $D^*D^*$                 \\\hline\hline
    $0^+$           & \SLJ{1}{S}{0}  & $-$                           & \SLJ{1}{S}{0},\SLJ{5}{D}{0}   \\\hline
    $1^+$           & $-$            & \SLJ{3}{S}{1},\SLJ{3}{D}{1}   & \SLJ{5}{D}{1}                 \\\hline
    $2^+$           & \SLJ{1}{D}{2}  & \SLJ{3}{D}{2}                 & \SLJ{5}{S}{2},\SLJ{1}{D}{2},\SLJ{5}{D}{2}, (\SLJ{5}{G}{2}) \\\hline
    $3^+$           & $-$            & \SLJ{3}{D}{3}, (\SLJ{3}{G}{3})  & \SLJ{5}{D}{3}, (\SLJ{5}{G}{3})   \\\hline
    $4^+$           & (\SLJ{1}{G}{4})  & (\SLJ{3}{G}{4})            & \SLJ{5}{D}{4}, (\SLJ{1}{G}{4},\SLJ{5}{G}{4},\SLJ{5}{I}{4})
\end{tabular}   
    }}
    \caption{Allowed $I=1$, $P=+$ partial waves $\SLJ{2S+1}{\ell}{J}$ for each $J \leq 4$ and $D^{(*)} D^{(*)}$ channel due to symmetry considerations.
    The waves in parentheses are not examined in this work.}
    \label{tab:partialwavebyJp}
\end{table}
\begin{table}[!t]
    \centering
    \setlength{\tabcolsep}{0.5em}{\renewcommand{\arraystretch}{1.5}{
        \begin{tabular}{c|c}
        Hadron        & $a_t m$         \\\hline\hline
        $\pi$         & $0.06906(13)$   \\\hline
        $D$           & $0.33281(9)$    \\\hline
        $D^*$       & $0.35464(14)$   \\\hline
        $D_0^*$     & $0.40170(18)$   \\\hline
    \end{tabular}
    \quad
    \begin{tabular}{c|c}
        Channel             & $a_t E_\thr$        \\\hline\hline
        $DD$               & $0.66562(13)$     \\\hline
        $DD^*$             & $0.68745(17)$     \\\hline
        $D^*D^*$           & $0.70928(20)$     \\\hline
        $DD^*_0$            & $0.73381(20)$     \\\hline
        $DD\pi$             & $0.73468(18)$     \\\hline
    \end{tabular}
    }}
    \caption{\label{tab:hadmasses}
    Relevant stable hadron masses (left) and thresholds (right) in lattice units, together with their statistical uncertainties~\cite{Dudek:2012gj,Moir:2016srx,Wilson:2023anv}.
    }
\end{table}

We use the same lattice setup and methodology as in Ref.~\cite{Whyte:2024ihh}, and here only briefly review key aspects and point out the differences.
Results were extracted from lattices generated from an anisotropic Wilson-Clover fermion action with $2+1$ flavors of dynamical quarks at three volumes, $L/a_s=16,20,24$, where the spatial lattice spacing is $a_s\approx 0.12$~fm~\cite{Edwards:2008ja}.
Those lattices are tuned to have hadron masses $m_{\pi} \approx 391\mev$, $m_D \approx 1886\mev$ and $m_{D^*} \approx 2010\mev$, such that the $D^*$ meson is below the $D\pi$ threshold and thus stable~\cite{Edwards:2012fx}.
The anisotropy ${\xi = a_s/a_t = 3.444(50)}$ was determined by fitting various stable hadron's energies to the relativistic dispersion relation, as detailed in Ref.~\cite{Wilson:2023anv}.
The temporal lattice spacing $a_t \approx \left( 5667\mev\right)^{-1}$ was previously determined using the $\Omega$ baryon mass and is used here to quote quantities in physical units~\cite{Edwards:2012fx}.
Other pertinent information on relevant masses and thresholds is shown in Table~\ref{tab:hadmasses}.

We compute matrices of lattice correlation functions ${\langle 0 \vert \mathcal{O}_i(t+t_\mathrm{src}) \mathcal{O}_j^\dagger(t_\mathrm{src}) \vert 0 \rangle}$ using the same distillation setup as in Ref.~\cite{Whyte:2024ihh}.
In particular, in this work the number of time sources $t_\mathrm{src}$ is $N_\mathrm{tsrc}=8$ for the $L/a_s=16,20$ lattices, and $N_\mathrm{tsrc}=4$ for the $L/a_s=24$ one.
The correlation functions are averaged over $t_\mathrm{src}$ for a higher statistical precision.
For each channel and volume, the finite-volume spectrum is then extracted up to the $DD\pi$ threshold using the variational method~\cite{Blossier:2009kd,Michael:1985ne,Luscher:1990ck}.
This is implemented by solving an associated generalized eigenvalue problem (GEVP) on each time slice~\cite{Dudek:2010wm}, yielding the resulting eigenvalues (principal correlators) $\lambda_{\mathfrak{n}}(t,t_0)$.
The lattice energies $E_{\mathfrak{n}}$ are obtained from fits to the double-exponential form
\begin{equation}
    \lambda_{\mathfrak{n}}(t,t_0) = (1-A_{\mathfrak{n}})e^{-E_{\mathfrak{n}}(t-t_0)} + A_{\mathfrak{n}} e^{-E_{\mathfrak{n}}^{\prime}(t-t_0)},
\end{equation}
where the second term absorbs residual contaminations from excited states.
For further interpreting the finite-volume spectrum, it is also useful to compute the operator-state overlaps $Z^\mathfrak{n}_i = \langle \mathfrak{n} | \mathcal{O}_i^\dagger(0) | 0 \rangle$ from the GEVP eigenvectors, as in Ref.~\cite{Whyte:2024ihh}.

Due to the breaking of infinite-volume rotation symmetry on the finite cubic lattice, the angular momentum $J$ is no longer a good quantum number.
The operators $\{\mathcal{O}_i\}$ thus need to be subduced into cubic symmetry irreps $\Lambda$~\cite{Johnson:1982yq,Dudek:2010wm}.
Parity is still a good quantum number in rest-frame irreps.
In this work, the operator bases are formed exclusively of meson-meson operators of the type~\cite{Dudek:2012gj}
\begin{equation}
    \mathcal{O}^{\dag}_{D^{(*)} D^{(*)}}(\vec{P}) = \sum_{\vec{p_1},\vec{p_2}} \left[ \mathrm{CGs} \right] \ \Omega_{D^{(*)}}^{\dag}(\vec{p_1}) \ \Omega_{D^{(*)}}^{\dag}(\vec{p_2})    \,, 
    \label{eq:memeopsimple}
\end{equation}
where $\vec{P}=\vec{p}_1+\vec{p}_2$ and the `$\mathrm{CGs}$' contain the necessary lattice Clebsch-Gordan coefficients to project the two-hadron interpolator into a given cubic symmetry irrep.
The optimized single-hadron operators $\Omega_{D^{(*)}}$ are constructed as described in Ref.~\cite{Dudek:2012gj}.
We include operators of the type shown in Eq.~\eqref{eq:memeopsimple} at overall zero spatial momentum and single-hadron momenta $\vec{p_1}^2 , \vec{p_2}^2 \le 4 \, (2\pi/L)^2$.
Using the same  $L/a_s=16$ lattice and methods as here, Ref.~\cite{Cheung:2017tnt} verified that the inclusion of compact tetraquark operators had no significant impact on the spectrum determination in the $I=0$ doubly charmed sector\footnote{Given that the tetraquark operator was not observed to influence the spectrum on the strongly attractive $I=0$ channel, one might anticipate it is also unimportant in $I=1$.}.
Following that observation, we do not employ tetraquark operators in this work.
The full list of operators used for each lattice volume is given in Appendix~\ref{apx:operatortables}. 

\begin{figure*}[!t]
    \centering
    \includegraphics[width=0.45\textwidth, trim={2.5mm 3mm 2.5mm 1mm},clip]{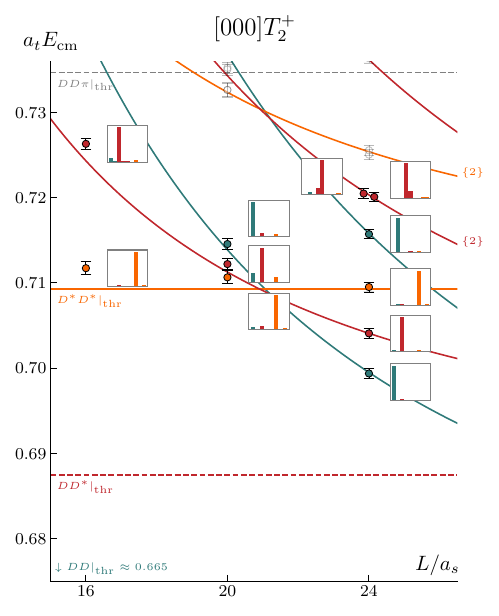}
    \includegraphics[width=0.45\textwidth, trim={2.5mm 3mm 2.5mm 1mm},clip]{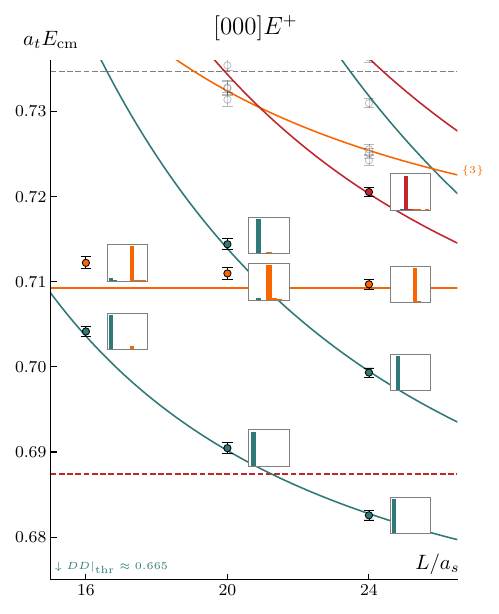}
    
    \caption{\label{fig:fvspectra2}
    Same as Fig.~\ref{fig:fvspectra1}, but for the $[000]T_2^+$ and $[000]E^+$ irreps.
    }
\end{figure*}

\section{\label{sec:fvspectra} Finite-volume spectra}

Here, the finite-volume lattice energies in the $I=1$ doubly charmed sector are presented.
The different cubic irreps are labelled as $[n_x n_y n_z] \Lambda^{P}$, with parity $P$ and total momentum $\vec{P} = \frac {2\pi}{L} (n_x, n_y, n_z)$.\footnote{As $\vec{P} = [000]$ is the only case here, we shall at times drop the $[000]$ from the the irrep notation.}
The subduction pattern of the partial waves from Table~\ref{tab:partialwavebyJp} into these irreps is shown in Table~\ref{tab:subductions}.
In Figs.~\ref{fig:fvspectra1} and \ref{fig:fvspectra2}, the computed energies are shown together with the relevant noninteracting energies and thresholds.
In Appendix~\ref{apx:negparity}, the negative parity $[000]A_1^-$, $[000]T_1^-$ and $[000]E^-$ irreps are briefly presented.

As noted by previous calculations on this lattice, unaccounted systematics, such as discretization effects, are present on the determination of finite-volume energies.
One manifestation of this is the mismatch between the $D$ and $D^*$ lattice energies and the continuum dispersion relation as shown Ref.~\cite{Wilson:2023anv}. We follow Ref.~\cite{Whyte:2024ihh} and adopt the conservative choice of an overall $a_t \delta_{syst} = 5 \times 10^{-4}$ systematic error added in quadrature to the statistical errors of the finite-volume lattice energies.
The error bars on the energies in Figs.~\ref{fig:fvspectra1} and \ref{fig:fvspectra2} correspond to this total uncertainty.

\begin{table}[!b]
    \centering
    \setlength{\tabcolsep}{0.5em}{\renewcommand{\arraystretch}{1.5}{
    \begin{tabular}{c||c|c|c}
$[n_x n_y n_z] \Lambda^{(P)}$   & $DD$           & $DD^*$         & $D^*D^*$         \\\hline\hline
$[000]A_1^+$                         & $\SLJ{1}{S}{0}$ & $-$             & $\SLJ{1}{S}{0}$, \SLJ{5}{D}{0}, $\SLJ{5}{D}{4}$   \\\hline
$[000]T_1^+$                         & $-$             & $\SLJ{3}{S}{1}, \SLJ{3}{D}{1}, \SLJ{3}{D}{3}$                & $\SLJ{5}{D}{1}$, $\SLJ{5}{D}{3}$, ($\SLJ{5}{D}{4}$)                 \\\hline
$[000]A_2^+$                         & $-$             & $\SLJ{3}{D}{3}$ & $\SLJ{5}{D}{3}$                  \\\hline
$[000]T_2^+$                         & $\SLJ{1}{D}{2}$ & $\SLJ{3}{D}{2}$, $\SLJ{3}{D}{3}$ & $\SLJ{5}{S}{2}$, ($\SLJ{1}{D}{2}$, $\SLJ{5}{D}{2,3,4}$)                   \\\hline
$[000]E^+$                           & $\SLJ{1}{D}{2}$ & $\SLJ{3}{D}{2}$ & $\SLJ{5}{S}{2}$, ($\SLJ{1}{D}{2}$, $\SLJ{5}{D}{2,4}$)                  \\
\end{tabular}
    }}
    \caption{
    Contribution of $I=1$, $D^{(*)}D^{(*)}$ partial waves to cubic symmetry irreps with positive parity ($P=+$), up to $\ell=2$.
    The waves are denoted by the spectroscopic notation $\SLJ{2s+1}{\ell}{J}$.
    The waves in parentheses are not considered in the scattering analysis involving those respective irreps, as will be detailed in Sec.~\ref{sec:scattering}.}
    \label{tab:subductions}
\end{table}

The operator-state overlaps $Z_i^\mathfrak{n}$ for each volume and irrep are given by the histograms in Figs.~\ref{fig:fvspectra1} and \ref{fig:fvspectra2}.
For each $i$, they are normalized by the value of $Z_i^\mathfrak{n}$ with the largest magnitude over all $\mathfrak{n}$.
The order of the operators within the histograms is given first by the opening of the associated threshold, and then by increasing associated noninteracting energy, corresponding to the ordering in Appendix~\ref{apx:operatortables}.
It is noted that all lattice energies are dominated by essentially one single meson-meson operator each, with the exception of the ones created by the operators degenerate in the limit of no hadron-hadron interactions, signaled by the number in curly brackets in Figs.~\ref{fig:fvspectra1} and \ref{fig:fvspectra2} and Tables~\ref{tab:operatortable0p},~\ref{tab:operatortable1p} and \ref{tab:operatortable2p}.
\footnote{Note that operator overlaps can depend on the details of the operator basis construction.
Here, the overlaps are only used as a qualitative guide to interpret the finite-volume spectrum.}
Across all irreps, the energies closely follow the noninteracting curves corresponding to their dominant operator content and are overall slightly shifted upwards in relation to the latter, indicating weakly repulsive interactions.
This is also the case near the crossing of noninteracting hadron-hadron energies, suggesting at most small coupled-channel effects.
Furthermore, small but non-negligible $S$-wave interactions are suggested by the relatively more pronounced energy shifts next to the $DD$, $DD^*$ and $D^*D^*$ thresholds in the $[000]A_1^+$, $[000]T_1^+$, and $[000]T_2^+, [000]E^+$ irreps, respectively.
Below threshold, the absence of energy levels indicates no bound states are present in the corresponding channels.
A quantitative study of these aspects is reserved to the next section.  

In the next section, the energy levels from the irreps presented above are used to constrain various scattering amplitudes parametrizations.
In particular, the $A_1^+$, $T_1^+$ and $T_2^+$, $E^+$ levels are used to constrain scattering in the $J^P=0^+, 1^+$ and $2^+$ channels, respectively.

\section{\label{sec:scattering} Scattering analysis}

We use the Lüscher method~\cite{Luscher:1990ux} and its extensions~\cite{Rummukainen:1995vs,Bedaque:2004kc,Kim:2005gf,He:2005ey,Lage:2009zv,Fu:2011xz,Leskovec:2012gb,Gockeler:2012yj,Hansen:2012tf,Briceno:2012yi,Guo:2012hv,Briceno:2014oea} to extract scattering amplitude parametrizations from the finite-volume lattice spectra energies.
In this method, the so-called quantization condition
\begin{equation}
    \mathrm{det} \ [{\bf 1}+ i{\pmb{\rho}} \cdot \bm{t} \cdot ({\bf 1} + i \pmb{ \cal M})] = 0
    \label{eq:luscher}
\end{equation}
is satisfied at the finite-volume energies in a spatial volume $L^3$ with periodic boundary conditions, where $\bm{t}(\Ecm)$ is the infinite-volume $t$-matrix.
All bold quantities are matrices in channel and angular momentum spaces subduced into cubic irreps~\cite{Thomas:2011rh}.
The known function $\pmb{{\cal M}} \big(\Ecm, L \big)$ depends on the \CM energy and spatial volume.
The phase-space factor $\pmb{\rho}$ is a diagonal matrix with nonzero components equal to $2 k_a / \Ecm$ on each two-hadron channel $a$, where $k_a = |\vec k_a| $ is the magnitude of their scattering \CM momentum.
Given $\bm{t}(\Ecm)$, the solutions of Eq.~\eqref{eq:luscher} give the finite-volume energies.
It is possible to determine $\bm{t}$ by minimizing a spectrum chi-squared ($\chi^2$) formed by the correlated differences of the lattice energies and the energies predicted by parametrizations of $\bm{t}$~\cite{Wilson:2014cna}.
The extracted parametrizations can be analytically continued into the complex-energy plane, where the presence or not of states is implied by pole singularities.
It is also important to check that the extracted parametrizations do not have physical-sheet poles with nonzero imaginary parts, which imply a violation of causality and are thus unphysical. 

\begin{figure*}[!t]
    \centering
    \includegraphics[width=\textwidth]{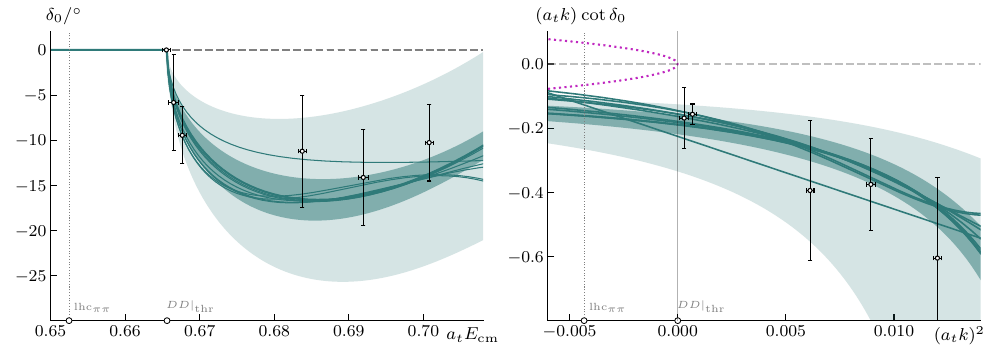}
    \caption{
    \label{fig:elastic0pbands}
    Phase shift (left) and $k \cot \delta_0$ (right) representations of the results for isovector $DD$ elastic scattering in $J^P=0^+$.
    The darker band represents the statistical error of the example parametrization quoted in Eq.~\eqref{eq:0pelasticpars}, while the lighter band is an envelope which accounts for the mass-anisotropy variation.
    The different curves correspond to the central value of all other reasonable fits listed in Table~\ref{tab:par_var_0pelastic}.
    The data points are obtained by using the lattice energies in Eq.~\eqref{eq:luscher}.
    The relevant thresholds and left-hand cut positions are indicated on the horizontal axis.
    The intersection of the $k \cot \delta_0$ curves with the purple dotted line indicates the position of $t$-matrix singularities on the corresponding parametrizations.
    }
\end{figure*}

Note that the finite-volume method used here, driven by Eq.~\eqref{eq:luscher}, does not take into account left-hand branch cuts~\cite{Green:2021qol}.
In addition, the potential effects of the latter are not explicitly included in the parametrizations extracted here.
The location of the left-hand cuts relevant for each of the channels analyzed is indicated within the following subsections.

When only the lowest hadron-hadron channel is open and a single partial wave is relevant, the scattering amplitude can be written in terms of a real phase shift $\delta_\ell$ and admits an \textit{effective range expansion}.
For $S$-wave scattering, this is given by
\begin{equation}
    k \cot \delta_0 = \frac 1a + \frac r2 k^2 + \mathcal{O}(k^4)  \,,
    \label{eq:ere}
\end{equation}
to first-order in $k^2$, which is related to the $S$-wave component of the $t$-matrix via $t = \left(k \cot \delta_0  - ik \right)^{-1} \Ecm/2$.
This $t$-matrix form implies an analytical continuation in $\Ecm$, and it can be shown that its singularities will in general correspond to solutions of $k \cot \delta_0 - ik = 0$ for complex $k$ (see Fig.~\ref{fig:elastic0pbands}).
In Sec.~\ref{sec:0pelastic}, a variation of the effective range expansion is considered, where an extra factor is included in the denominator of the expansion above.

In the energy region above the first two-hadron inelastic threshold or where multiple partial waves contribute, a suitable parametrization for the scattering amplitude is given by the so-called K-matrix
\begin{equation}
    [\bm{t}^{-1}]_{\ell a,\ell' b} = (2k_a)^{-\ell} [\Kmatb^{-1}]_{\ell a,\ell' b} (2k_b)^{-\ell'} + \delta_{\ell \ell'}I_{ab}
    \label{eq:tinv}
\end{equation}
for a given $J^P$, where the orbital angular momentum ($\ell,\ell'$) and channel ($a,b$) indices were spelled out together with the threshold factors $(2k_a)^{-\ell}$.
The intrinsic spin and total angular momentum indices are left implicit.
Through Eq.~\eqref{eq:tinv}, unitarity of the scattering S-matrix $\bm{S} = {\bf 1} + 2 i {\pmb{\rho}}^{1/2} \cdot \bm{t} \cdot {\pmb{\rho}}^{1/2}$ is automatically ensured given that the K-matrix is real and symmetric for real invariant mass $s$, and that $\Imag I_{ab} = 0$ below the associated threshold and $\Imag I_{ab} = -\delta_{ab} \rho_a$ above it.
The exact choice of the real part of $I_{ab}$ is arbitrary, and we employ both the ``simple'', given by $\Real I_{ab} = 0$, and the ``Chew-Mandelstam'' phase-space versions, the latter obtained from a dispersive integral~\cite{Wilson:2014cna}.
In particular, we implement K as an expansion in powers of $s$ about a given $s_\mathrm{ref}$, of the form
\begin{equation}
    \Kmat_{\ell SJa,\ell'S'Jb}(s) = \sum_{n} \gamma_{\ell SJa,\ell'S'Jb}^{(n)} \hat s^n,
    \label{eq:k_poly}
\end{equation}
where $\hat s \equiv s - s_\mathrm{ref}$, and the real coefficients $\gamma$ are determined by a fit to the spectrum, as described above.
The value $a_t^2 s_\mathrm{ref} = (0.70)^2$ is used throughout the following sections.\footnote{Even though the expansion point $s_\mathrm{ref}$ is not of physical significance, the choice of values closer to the lowest threshold led to smaller correlations between the resulting parameters in this work.}

In some cases, the amplitude parametrizations above can give rise to bound-state poles when extrapolated too far from the data, where no corresponding energy levels are present in the lattice calculation.
In any case, they should not be considered too far below threshold as their validity is typically limited by nearby singularities such as left-hand branch points.
When necessary, we use a procedure similar to the one in Ref.~\cite{Yeo:2024chk} to ensure that such spurious bound-state poles are not present near the region constrained by the data.
To do this, a term is added to the spectrum $\chi^2$ that smoothly increases with the proximity of these singularities to the determined spectrum, disfavoring these parameter values.
In particular, we choose the latter to be the region between the lowest two-hadron threshold and the first left-hand cut just below it.
This procedure is detailed in Appendix~\ref{apx:polepush}.
Note that we also check for physical-sheet poles with nonzero imaginary parts close to the region of the complex plane that is constrained by the energy levels.

In the following, we present the results of fits to the lattice spectrum below the $DD\pi$ threshold using the parametrizations and procedures outlined above.

\subsection{\label{sec:0p} $J^P=0^+$}

\subsubsection{\label{sec:0pelastic} Elastic $DD$ scattering below $D^*D^*$ threshold}

We start by considering elastic $DD$, $S$-wave scattering in the $J^P=0^+$ channel. For this, energy levels up to the the $D^*D^*$ kinematic threshold at $a_t \Ecm \approx 0.70$ are used, corresponding to the $6$ lowest points from $A_1^+$ in Fig.~\ref{fig:fvspectra1}.

Here, a relatively simple parametrization which reasonably describes the energy levels is the K-matrix with Chew-Mandelstam phase space and parameters
\begin{small}
    \begin{center}
    \renewcommand{\arraystretch}{1.4}
    \begin{tabular}{rll}
        $\gamma^{(0)} = $ & $(-0.554 \pm 0.084 \pm 0.434)$ & \multirow{2}{*}{ $\begin{bmatrix*}[r]   1.00 &  -0.18\\
        &  1.00\end{bmatrix*}$ } \\ 
        $\gamma^{(1)} = $ & $(21.2 \pm 4.9 \pm 1.8) \cdot a_t^2 $ & \\[1.3ex]
        &\multicolumn{2}{l}{ $\chi^2/ N_\mathrm{dof} = \frac{2.81}{6-2} = 0.70$\,,}
    \end{tabular}
    \begin{equation}
    \label{eq:0pelasticpars}
    \end{equation}
    \end{center}
\end{small}
where exclusively in this section, we use the shorthand notation ${\gamma^{(n)} \equiv \gamma^{(n)}_{DD \SLJc{1}{S}{0} \to DD \SLJc{1}{S}{0}}}$.
The first error is statistical, while the second represents the variation resulting from fits where the stable hadron masses and anisotropy were taken one-sigma away from their central values one at a time (see Table~\ref{tab:hadmasses}), referred to as mass-anisotropy variation.
The values in square brackets are the statistical correlations between the resulting parameters.
Any possible terms that are not listed had their coefficients set to zero.

To assess the dependence of the results on the K-matrix form used, we produce parametrizations variations which are summarized in Table~\ref{tab:par_var_0pelastic} in Appendix~\ref{apx:partables}.
Different K-matrix expansions are employed with terms up to second-order in $s$ and with different phase space choices.
It is observed that at least a linear term in $s$ is needed for a reasonable description of the energy levels.
We also consider effective range-type parametrizations, where it is readily noted that the scattering length only is not enough to reasonably describe the data.
The inclusion of the effective range parameter leads to ${a a_t^{-1} = -4.43 \pm 0.87}$ and ${r a_t^{-1} = -45.3 \pm 7.7}$, but with a final $\chi^2/N_\mathrm{dof} = 2.22$ after ensuring no spurious bound-state poles are present between the $DD$ threshold and the nearest left-hand cut.\footnote{For the pion mass used here, the scalar $\sigma$ state is bound~\cite{Briceno:2016mjc}, and thus leads to the left-hand branch point closest to the $DD$ threshold at $a_t E_{\sf lhc,\pi\pi} \approx 0.65$.As the effective range parametrization is the first instance where the spurious-pole procedure is employed, we detail this specific case in Appendix~\ref{apx:polepush} (see Fig.~\ref{fig:polexc_demo}).}
On the other hand, the use of a $k^4$-term in Eq.~\eqref{eq:ere} is free of spurious bound state poles but leads to physical-sheet poles with nonzero imaginary components.
The latter violate fundamental analytical properties expected from the scattering amplitude and thus such parametrizations are not further considered.

In Ref.~\cite{Romero-Lopez:2019qrt}, a expansion similar to the effective range but with an extra parameter $\kappa$ in the pole factor, namely $k \cot \delta(k) = (1/a + rk^2/2) (1-k^2/\kappa^2)^{-1}$, is considered as a possible way of extending the validity of the expansion in the presence of a (non-spurious) bound state.
This form first appeared in the context of phenomenological analyses of neutron-deuteron data~\cite{Reiner:1969mmv}.
We employ parametrizations of this kind with up to $k^2$-terms in the numerator, and observe that its application improves the compatibility with the data (see Table~\ref{tab:par_var_0pelastic}).
In particular, it does not introduce complex physical-sheet poles.
For reference, when this form is used with the effective range parameter fixed to zero, the scattering length results in $a = -0.191 \pm 0.034$~fm.
We stress that this value only includes the statistical uncertainty of a single parametrization, and does not take into account mass-anisotropy variations, scale setting uncertainties or other systematic uncertainties.
We note that typically extracted values for $(a_t \kappa)^2 \approx 0.02$ are not much larger than $(a_t k)^2$ in the region constrained by the data.
This indicates that this pole-like term cannot be represented by only a few terms of the regular effective range expansion.

In Table~\ref{tab:par_var_0pelastic}, all parametrizations and their respective reduced chi-squared values $\chi^2/N_\mathrm{dof}$ are presented.
In all variations shown, it was ensured that spurious bound-state poles do not appear near the energy region considered, in the way detailed in Appendix~\ref{apx:polepush}.

Fig.~\ref{fig:elastic0pbands} shows the phase-shift $\delta_0$ and $k \cot \delta_0$ from the parametrizations variations.
As indicated on the respective table, the ones yielding reasonable fits are represented by individual curves, while the bands correspond to the example parametrization in Eq.~\eqref{eq:0pelasticpars}.
The interactions are observed to be consistent with a weakly repulsive system, and no pole
singularities corresponding to states are found in the energy region considered below $D^*D^*$ threshold.

\subsubsection{Coupled-channel $DD - D^*D^*$ scattering}
\label{sec:0pcoupled}

To go higher in energy, we need to consider the dynamical coupling between $DD$ and $D^*D^*$ in $J^P=0^+$, where the $DD$, $S$ wave as well as $D^*D^*$, $S$ and $D$ waves are possible contributions. 
For that, a further $8$ energy levels in the $A_1^+$ irrep below the $DD\pi$ threshold are included in addition to the ones used in the elastic case. 
The violation of infinite-volume rotational symmetry is manifest as $J^P=4^+$, $D$- and $G$-wave contributions to $A_1^+, T_1^+, T_2^+$ and $E^+$ (see Table~\ref{tab:subductions}).
Here, we briefly consider some of these contributions to the $A_1^+$ levels but not to the $T_1^+$ or to the $T_2^+$ and $E^+$ ones, which are respectively discussed in Secs.~\ref{sec:1p} and ~\ref{sec:2p}.
Overall, all partial waves in $J^P=0^+$ are taken into account, while various $G$-waves ($\ell=4$) are the lowest ones ignored in $J^P=4^+$ (see Table~\ref{tab:partialwavebyJp}).

For the coupled-channel $J^P=0^+$ case, the example parametrization is chosen to be the K-matrix with simple phase space and parameters
\begin{widetext}
    \begin{small}
        \begin{center}
        \renewcommand{\arraystretch}{1.4}
        \begin{tabular}{rll}
            $\gamma^{(0)}_{DD \SLJc{1}{S}{0} \to DD \SLJc{1}{S}{0}} =$ & $(-0.584 \pm 0.084 \pm 0.427)$ & \multirow{4}{*}{ $\begin{bmatrix*}[r]   1.00 &   0.54 &   0.08 &  -0.62\\
            &  1.00 &   0.06 &  -0.52\\
            &&  1.00 &   0.01\\
            &&&  1.00\end{bmatrix*}$ } \\ 
            $\gamma^{(0)}_{DD \SLJc{1}{S}{0} \to D^*D^* \SLJc{1}{S}{0}} = $ & $(1.11 \pm 0.22 \pm 2.25)$ & \\
            $\gamma^{(0)}_{D^*D^* \SLJc{1}{S}{0} \to D^*D^* \SLJc{1}{S}{0}} = $ & $(0.59 \pm 0.16 \pm 0.17)$ & \\
            $\gamma^{(0)}_{DD \SLJc{1}{S}{0} \to D^*D^* \SLJc{5}{D}{0}}  = $ & $(32.8 \pm 3.6 \pm 2.2) \cdot a_t^2 $ & \\[1.3ex]
            &\multicolumn{2}{l}{ $\chi^2/ N_\mathrm{dof} = \frac{9.79}{14-4} = 0.98$\,,}
            \end{tabular}
        \begin{equation}
        \label{eq:0pcoupledpars}
        \end{equation}
        \end{center}
    \end{small}
\end{widetext}
where the uncertainties are defined as in Eq.~\eqref{eq:0pelasticpars}.
In addition to the $DD \SLJc{1}{S}{0}$ diagonal parameter, which has a similar value to the elastic case, the $D^*D^* \SLJc{1}{S}{0}$ diagonal parameter in Eq.~\eqref{eq:0pcoupledpars} is also found to be non zero.
The latter suggests $D^*D^*$, $S$-wave interactions in the coupled-channel region.

\begin{figure*}[!t]
    \centering
    \includegraphics[width=0.95\textwidth, trim={3mm 3mm 2mm 2mm},clip]{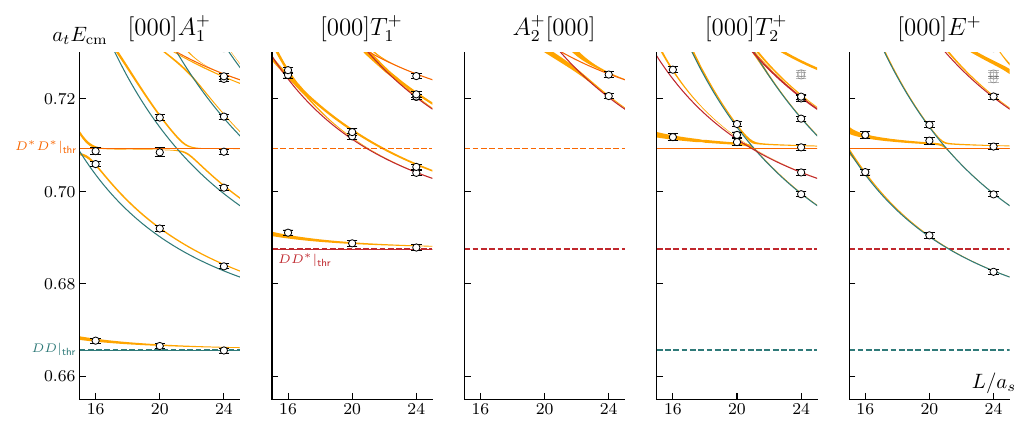}
    \caption{\label{fig:getfinite}
    Lattice energies (black points) and finite-volume spectrum as a function of $L/a_s$ (orange bands) on each irrep considered for analyzing $J^P=0^+,1^+$ and $2^+$ coupled-channel scattering.
    The spectrum bands are derived from the example parametrizations quoted in Eqs.~\eqref{eq:0pcoupledpars}, \eqref{eq:1p3pcoupledpars}, \eqref{eq:2pcoupledpars}.
    The bands only account for the statistical uncertainties on the quoted parameters.
    Non-interacting energies are represented by solid lines and threshold by dashed lines, following the same color scheme of Figs.~\ref{fig:fvspectra1} and \ref{fig:fvspectra2}.
    }
\end{figure*}

The left-most panel of Fig.~\ref{fig:getfinite} shows the finite-volume spectrum obtained from the solution of the quantization condition Eq.~\eqref{eq:luscher} using the example parametrization in Eq.~\eqref{eq:0pcoupledpars}.
Those energies and the associated statistical error bands describe reasonably well the lattice data, and reproduce the small shifts in relation to the nearby noninteracting $DD$ and $D^*D^*$ levels.

To further assess the consistency of the results above, several K-matrix parametrization variations are obtained by allowing for different terms up to linear order in $s$, as detailed in Table~\ref{tab:par_var_0pcoupled}.
When necessary, spurious bound states appearing in these variations are ensured to be far enough from the region constrained by the data (see Appendix~\ref{apx:polepush}).

\begin{figure*}[!t]
    \centering
    \includegraphics[width=0.49\textwidth, trim={2mm 1mm 0mm 0.5mm},clip]{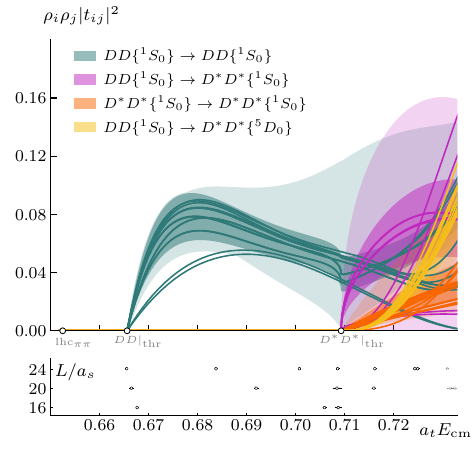}
    \includegraphics[width=0.49\textwidth, trim={2mm 1mm 0mm 0.5mm},clip]{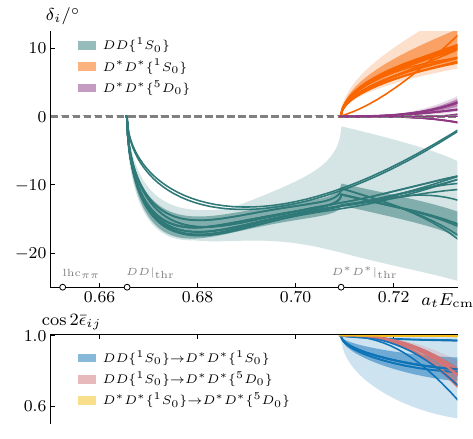}

    \caption{\label{fig:coupled_0p}
    Left: Normalized squared amplitudes (upper) and lattice energies (lower) for coupled-channel $DD, D^*D^*$ scattering in $I=1$ and $J^P=0^+$.
    Other amplitudes that are consistent with zero are not shown for clarity.
    Right: Phase shift (upper) and inelasticity (lower) from the generalized Stapp parametrization applied to the same system, as described in the text.
    The inner and outer bands correspond to the statistical error and mass-anisotropy envelope of the example parametrization, as described in the text and quoted in Eq.~\eqref{eq:0pcoupledpars}.
    The curves represent the central value result of all parametrization variations yielding reasonable fits, as detailed in Table~\ref{tab:par_var_0pcoupled}.
    }
\end{figure*}

\clearpage

\begin{figure*}[!t]
    \centering
    \includegraphics[width=0.49\textwidth, trim={2mm 2mm 0mm 0.5mm},clip]{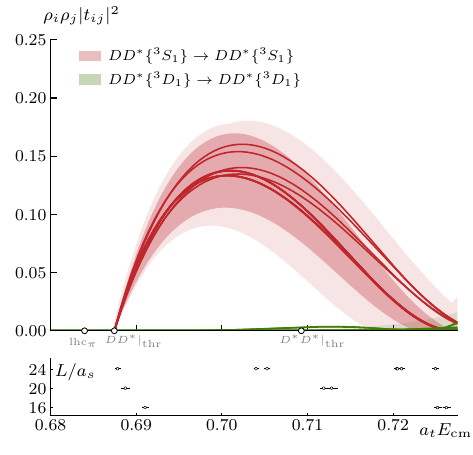}
    \includegraphics[width=0.49\textwidth, trim={2mm 2mm 0mm 0.5mm},clip]{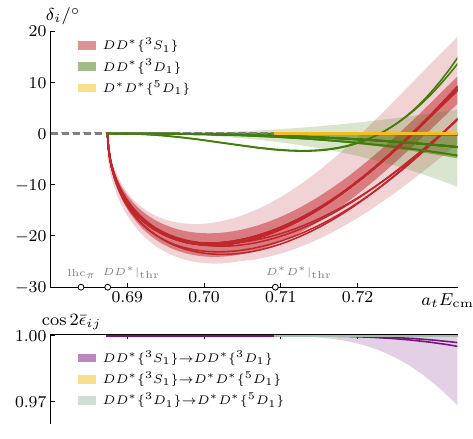}
    \caption{\label{fig:coupled_1p3p}
    The same as Fig.~\ref{fig:coupled_0p}, but for the coupled-channel $DD^* - D^*D^*$ in $J^P=1^+$.
    The example parametrizations shown here as bands are also quoted in Eqs.~\eqref{eq:1p3pcoupledpars}.
    The individual curves represent other parametrization variations yielding reasonable descriptions of the data, detailed in Table~\ref{tab:par_var_1p3pcoupled}.
    }
\end{figure*}

It is noted that at least one nonzero diagonal $D^*D^* \SLJc{1}{S}{0}$ term is needed to describe the three energies near the $D^*D^*$ threshold and thus this is included in all parametrization variations.
It is also verified that, among the partial waves considered, either a nonzero ${DD \SLJc{1}{S}{0} \to D^*D^* \SLJc{5}{D}{0}}$ or a diagonal $D^*D^* \SLJc{5}{D}{4}$ K-matrix parameter is required to reproduce one of the two near-degenerate energies at $a_t \Ecm \approx 0.725$ in the largest volume.
We further observe that an off-diagonal ${D^*D^* \SLJc{1}{S}{0} \to D^*D^* \SLJc{5}{D}{0}}$ K-matrix parameter is consistent with zero.
These observations indicate that the $D$-wave contributions from either $J^P=0^+$ or $J^P=4^+$ are small and constrained to be consistent with zero in the region considered, even though at least some of them are necessary to reproduce all the levels computed on the lattice.
Finally, a parameter allowing for the coupling of $DD$ to $D^*D^*$ via an $S$ wave is included in some parametrizations but it is overall consistent with zero, indicating these channels are weakly coupled.

In the left panel of Fig.~\ref{fig:coupled_0p}, the squared amplitudes from the example parametrization are represented by bands, while the variations are depicted by individual curves.
The size of the $DD \SLJc{1}{S}{0} \to DD \SLJc{1}{S}{0}$ squared-amplitude in the constrained region is consistent with weak $DD$ interactions.
Taking uncertainties into account, this wave is observed to be the largest in magnitude when compared to all other waves.
The $DD \SLJc{1}{S}{0} \to D^*D^* \SLJc{1}{S}{0}$ amplitude has a more abrupt rise in the parametrizations with only constant terms but is consistent with zero when taking into account all variations, indicating this is small and not well constrained by the data in the energy region probed here.
No resonances or bound states are found either below or above the $D^*D^*$ threshold in the energy region considered.
Here and in all analogous figures in this work, some amplitudes that are very small or consistent with zero are not shown for clarity.

We also consider the generalized Stapp parametrization to visualize the scattering amplitude results, as introduced in Ref.~\cite{Woss:2019hse}.
In this, the unitary S-matrix is minimally presented in terms of phases shifts $\delta_i$ and mixing angles $\bar \epsilon_{ij}$, where $i$ and $j$ denote the partial-wave channels considered.
The usual elastic definition of the phase shifts is recovered in the limit of vanishing channel coupling.
Instead of $\bar \epsilon_{ij}$, it is useful to show $\cos 2 \bar \epsilon_{ij}$ as it does not depend on an arbitrary phasing of the hadron states, and is similar to the so-called inelasticity $\eta$ used to describe two-channel systems~\cite{Briceno:2017qmb}.
The right panel of Fig.~\ref{fig:coupled_0p} shows the Stapp parametrization for the amplitudes considered here.
The $DD \SLJc{1}{S}{0}$ phases shift shown in the upper panel can then be readily compared below the $D^*D^*$ threshold to the elastic case in Fig.~\ref{fig:elastic0pbands}.
Above threshold, one observes the turn on of the $D^*D^* \SLJc{1}{S}{0}$ $S$-wave phase shift to positive values, a signature of attractive interactions.
In the lower panel, the values of the inelasticity $\cos 2 \bar \epsilon_{ij}$ between the $DD$, $S$ wave and the other two partial-wave channels deviate slightly from $1$, considering the large uncertainties.

\subsection{$J^P=1^+$: $DD^* - D^*D^*$ scattering}
\label{sec:1p}

We now consider $DD^*$ and $D^*D^*$ scattering in the isovector ${J^P=1^+}$ channel.
In this case, $DD^*$ scattering appears as $S$ and $D$ waves, and an additional $D^*D^*$, $D$ wave above the corresponding threshold.
We remind the reader that the partial waves allowed here are quite different from $I=0$ ones~\cite{Whyte:2024ihh}.
Because the $T_1^+$ irrep contains contributions from ${J^P=3^+}$ as well as from ${J^P=1^+}$ (see Table~\ref{tab:subductions}), we also consider $A_2^+$ to further resolve the former.
Note that there are ${J^P=3^+}$ contributions to ${T^+_2}$, but we do not consider that irrep in this section.
The amplitude parametrizations presented in the following are thus extracted from fits to the $15$ finite-volume energies across the $T_1^+$ and $A_2^+$ irreps (see Fig.~\ref{fig:fvspectra1}).
Overall, the lowest partial wave neglected in this analysis is the $D^*D^*\SLJc{5}{D}{4}$ one, arising from partial-wave mixing, while all $J^P=1^+$ waves are taken into account.

Initially, in order to provide a more direct comparison to Refs.~\cite{Chen:2022vpo,Meng:2024kkp}, we extract elastic $DD^*$ effective range-type parametrizations employing only a subset of the $[000]T_1^+$ levels.
The scattering length by itself can reasonably describe only the lowest three energies near the $DD^*$ threshold.
If one assumes that coupled-channel effects are small and consider further three energies containing $S$-wave contributions below $DD\pi$,\footnote{In Fig.~\ref{fig:fvspectra1}, these correspond to the upper points from the near-degenerate pairs next to the lowest noninteracting curve indicated by ``$\{2\}$'' across all volumes in the $[000]T_1^+$ irrep.} then adding an extra parameter is needed to achieve a reasonable description of the data (see Table~\ref{tab:par_var_1pelastic}).
The parametrization where the scattering length is modified by a pole term results in \text{$a = -0.316 \pm 0.042$}~fm, and $(a_t \kappa)^2 \approx -0.015$.
The value quoted in physical units is taken from a single parametrization, and it does not include mass-anisotropy variations, scale setting uncertainties or other systematic uncertainties.
Further comments are left to Sec.~\ref{sec:conclusions}.

To go higher into the coupled-channel region, we take the example parametrization to be
\begin{widetext}
    \begin{small}
        \begin{center}
        \renewcommand{\arraystretch}{1.4}
        \begin{tabular}{rll}
            $\gamma^{(0)}_{DD^* \SLJc{3}{S}{1} \to DD^* \SLJc{3}{S}{1}} = $ & $(-1.50 \pm 0.20 \pm 0.21)$ & \multirow{6}{*}{ $\begin{bmatrix*}[r]   1.00 &  -0.68 &   0.00 &   0.36 &   0.49 &  -0.02\\
            &  1.00 &  -0.01 &  -0.04 &  -0.14 &   0.05\\
            &&  1.00 &   0.00 &   0.02 &   0.00\\
            &&&  1.00 &   0.09 &  -0.33\\
            &&&&  1.00 &  -0.28\\
            &&&&&  1.00\end{bmatrix*}$ } \\ 
            $\gamma^{(1)}_{DD^* \SLJc{3}{S}{1} \to DD^* \SLJc{3}{S}{1}} =   $ & $(52.1 \pm 5.3 \pm 7.3)$ & \\
            $\gamma^{(1)}_{DD^* \SLJc{3}{D}{1} \to DD^* \SLJc{3}{S}{1}} =   $ & $(0 \pm 213 \pm 0) \cdot a_t^4$ & \\
            $\gamma^{(0)}_{DD^* \SLJc{3}{D}{1} \to DD^* \SLJc{3}{D}{1}} =   $ & $(-31 \pm 26 \pm 65) \cdot a_t^4$ & \\
            $\gamma^{(0)}_{DD^* \SLJc{3}{D}{3} \to DD^* \SLJc{3}{D}{3}} =   $ & $(-71 \pm 20 \pm 87) \cdot a_t^4$ & \\
            $\gamma^{(0)}_{DD^* \SLJc{3}{D}{3} \to D^*D^* \SLJc{5}{D}{3}} = $ & $(-357 \pm 111 \pm 150) \cdot a_t^4$ & \\[1.3ex]
            &\multicolumn{2}{l}{ $\chi^2/ N_\mathrm{dof} = \frac{9.07}{15-6} = 1.01$\,,}
        \end{tabular}
        \begin{equation}
        \label{eq:1p3pcoupledpars}
        \end{equation}
        \end{center}
    \end{small}
\end{widetext}
where the Chew-Mandelstam phase space was used.
The uncertainties correspond to statistical errors and the mass-anisotropy variations, as in Eq.~\eqref{eq:0pelasticpars}.\footnote{We show the $DD^* \SLJc{3}{D}{1} \to DD^* \SLJc{3}{S}{1}$ parameter that is statistically consistent with zero to demonstrate the very weak coupling between those two partial waves. The very small mass-anisotropy variations compared to the statistical uncertainty results in the vanishing of the right-most uncertainty to the quoted precision.}
In this example, it is notable the dominance of the diagonal $DD^* \SLJc{3}{S}{1}$ terms, and the parameter coupling $DD^* \SLJc{3}{S}{1}$ to $DD^* \SLJc{3}{D}{1}$ is consistent zero.
For this and all the other coupled-channel parametrizations in this section, no spurious bound-state poles are present near the energy region considered.

The $[000]T^+_1$ and $[000]A^+_2$ panels in Fig.~\ref{fig:getfinite} show the solutions of the quantization condition using the example parametrization Eq.~\eqref{eq:1p3pcoupledpars}.
The lattice data, including the near-degenerate levels mentioned above, are reasonably reproduced by the statistical bands, including the small upward shift in relation to the noninteracting energies.

\begin{figure}[!t]
    \centering
    \includegraphics[width=0.45\textwidth, trim={2mm 2mm 0mm 0.5mm},clip]{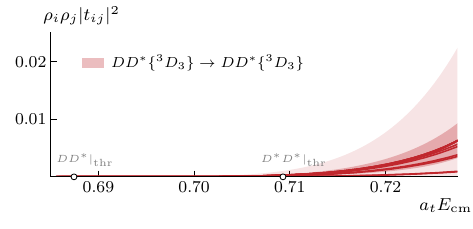}
    \caption{\label{fig:coupled_3p}
    Same as the left upper panel Fig.~\ref{fig:coupled_1p3p}, but the for $DD^*\SLJc{3}{D}{3}$ partial wave extracted in conjunction to the ${J^P=1^+}$ ones shown in Fig.~\ref{fig:coupled_1p3p}.
    The corresponding parameter is also quoted in Eq.~\eqref{eq:1p3pcoupledpars}.
    }
\end{figure}

Several parametrization variations are obtained for coupled-channel scattering using K-matrices up to linear order in $s$, as listed in Table~\ref{tab:par_var_1p3pcoupled}.
Throughout all these, a non-zero parameter with a $DD^* \SLJc{3}{S}{1}$ contribution is necessary to reproduce the levels considered via the quantization condition Eq.~\eqref{eq:luscher}.
Furthermore, it appears that both constant and linear-order  $DD^* \SLJc{3}{S}{1}$ diagonal terms are needed to achieve reasonable $\chi^2/ N_\mathrm{dof}$ in all variations.
Even though necessary to reproduce all levels using the quantization condition, all $D$-wave terms in $J^P=1^+$ are consistent with zero on all parametrization variations.
Only constant terms are used in the parametrization of the $J^P=3^+$ amplitudes.
The nonzero diagonal $DD^* \SLJc{3}{D}{3}$ K-matrix parameter was required to reproduce the near-degenerate levels at $a_t \Ecm \approx 0.720$ and $a_t \Ecm \approx 0.725$ in the largest volume (see Fig.~\ref{fig:fvspectra1}).

The left panel of Fig.~\ref{fig:coupled_1p3p} shows the squared amplitudes of the example parametrization in Eq.~\eqref{eq:1p3pcoupledpars} with its statistical error band and mass-anisotropy envelope, together with individual curves corresponding to parametrization variations.
These imply relatively weak interactions in the isovector $DD^*, J^P=1^+$ channel, which is also compatible with the results from other lattice works realized only in the elastic region~\cite{Chen:2022vpo,Meng:2024kkp}.
The dominant contribution comes from the ${DD^* \SLJc{3}{S}{1} \to DD^* \SLJc{3}{S}{1}}$ partial waves. 
No poles corresponding to resonances or bound states are found in this channel in the energy region considered.

In the right panel of Fig.~\ref{fig:coupled_1p3p}, the sign of the $DD^*$, $S$-wave phase shift in the Stapp parametrization shows that this interaction is repulsive, while for the $D$-waves the results are consistent with zero.
Additionally, the vanishing mixing angles, and thus inelasticities ${\cos 2 \bar \epsilon_{ij} \approx 1}$, demonstrate the very weak coupling between the partial-wave channels considered here.

Fig.~\ref{fig:coupled_3p} shows the only $J^P=3^+$ amplitude included in all parametrization variations, namely $DD^*\SLJc{3}{D}{3}$, which was necessary to reproduce some of the energy levels.
All the other $J^P=3^+$ contributions included in some parametrization variations have similar or smaller magnitude than $DD^*\SLJc{3}{D}{3}$.

\subsection{\label{sec:2p} $J^P=2^+$: $DD - DD^* - D^*D^*$ scattering}

We close the analysis section by presenting the results for coupled $DD, DD^*, D^*D^*$ scattering in $J^P=2^+$.
In this case, the lowest partial wave is $D^*D^*\SLJc{5}{S}{2}$ followed by various $D$-waves in $J^P=2^+$.
We use the spectra in the $T_2^+$ and $E^+$ irreps, which contain contributions from $J^P=2^+$, but also from $J^P=3^+,4^+$ due to partial-wave mixing.
In total, $19$ energy levels distributed across the $T_2^+$ and $E^+$ irreps are used to constrain the relevant amplitudes.

Several $D^*D^*$, $D$ waves in $J^P=2^+,3^+$ and $4^+$ are possible contributions to the $T_2^+$ and $E^+$ irreps, as listed in Table~\ref{tab:subductions}.
Note however that a $S$ wave is also present in the $D^*D^*$ channel, which is expected to dominate due to the near-threshold suppression of the $D$ waves.
We check the latter are only necessary to reproduce the near-degenerate energies just below the $DD\pi$ threshold in the largest volume.\footnote{These correspond to the faint points just above the opaque ones for the $L/a_s=24$, $[000]T_2^+$ and $[000]E^+$ irreps in Fig.~\ref{fig:fvspectra2}.}
For these reasons, we do not include those energies and neglect the $D^*D^*$, $D$ waves to extract the other contributions in the following.

The example parametrization is taken as
\begin{widetext}
    \begin{small}
        \begin{center}
        \renewcommand{\arraystretch}{1.4}
        \begin{tabular}{rll}
            $\gamma^{(0)}_{D^*D^* \SLJc{5}{S}{2} \to D^*D^* \SLJc{5}{S}{2}} = $ & $(-2.61 \pm 0.39 \pm 0.43)$ & \multirow{5}{*}{ $\begin{bmatrix*}[r]   1.00 &   0.36 &   0.42 &   0.42 &   0.23\\
            &  1.00 &   0.46 &   0.09 &  -0.49\\
            &&  1.00 &   0.41 &   0.13\\
            &&&  1.00 &   0.06\\
            &&&&  1.00\end{bmatrix*}$ } \\ 
            $\gamma^{(0)}_{DD^* \SLJc{3}{D}{2} \to DD^* \SLJc{3}{D}{2}} = $  & $(-57 \pm 21 \pm 78) \cdot a_t^4$ & \\
            $\gamma^{(0)}_{DD \SLJc{1}{D}{2} \to DD \SLJc{1}{D}{2}} = $ & $(-12 \pm 11 \pm 54) \cdot a_t^4$ & \\
            $\gamma^{(0)}_{DD \SLJc{1}{D}{2} \to D^*D^* \SLJc{5}{S}{2}} = $ & $(6.7 \pm 3.7 \pm 14.2) \cdot a_t^2$ & \\
            $\gamma^{(0)}_{DD^* \SLJc{3}{D}{3} \to DD^* \SLJc{3}{D}{3}} = $ & $(-99 \pm 31 \pm 30) \cdot a_t^4$ & \\[1.3ex]
            &\multicolumn{2}{l}{ $\chi^2/ N_\mathrm{dof} = \frac{15.53}{20-5} = 1.03$\,,}
            \end{tabular}
        \begin{equation}
        \label{eq:2pcoupledpars}
        \end{equation}
        \end{center}
    \end{small}
\end{widetext}
where the Chew-Mandelstam phase space was used.
The most notable feature of this parametrization is a nonzero ${D^*D^* \SLJc{5}{S}{2}}$ diagonal K-matrix parameter, which is needed to account for the energy shifts next to the $D^*D^*$ threshold.

\begin{figure*}[!t]
    \centering
    \includegraphics[width=0.49\textwidth, trim={2mm 2mm 0mm 0.5mm},clip]{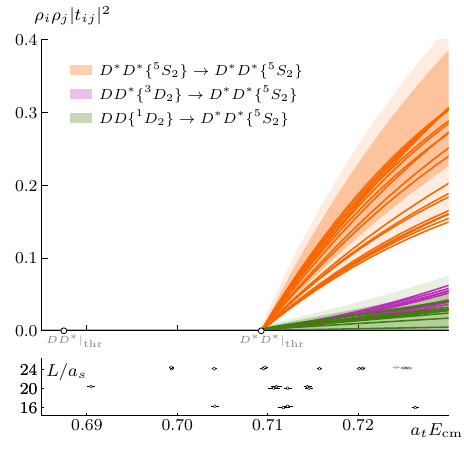}
    \includegraphics[width=0.49\textwidth, trim={2mm 2mm 0mm 0.5mm},clip]{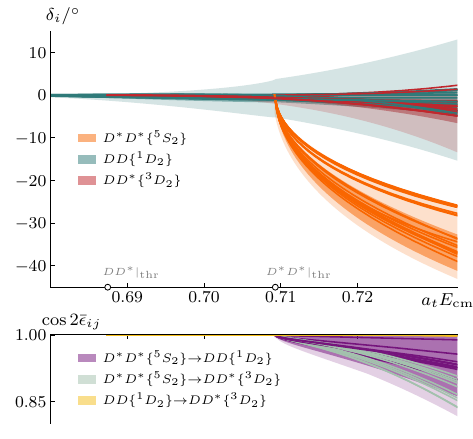}
    \caption{\label{fig:coupled_2p}
    Same as Fig.~\ref{fig:coupled_0p}, but the for coupled-channel $DD - DD^* - D^*D^*$ in $J^P=2^+$.
    The example parametrizations shown here as bands are also quoted in Eqs.~\eqref{eq:2pcoupledpars}.
    The individual curves represent other parametrization variations yielding reasonable descriptions of the data, detailed in Table~\ref{tab:par_var_2pcoupled}.
    }
\end{figure*}

The $[000]T^+_2$ and $[000]E^+$ panels in Fig.~\ref{fig:getfinite} show the spectrum resulting from the example parametrization Eq.~\eqref{eq:0pcoupledpars} via the quantization condition, where the lattice data is observed to be well described.
The upward shift to noninteracting levels near the $D^*D^*$ threshold is reproduced by the example parametrization.

Several variations of the K-matrix parametrization are considered up to linear order in $s$, as detailed in Table~\ref{tab:par_var_2pcoupled}.
When necessary, the procedure of Appendix~\ref{apx:polepush} is used to ensure that no spurious bound-states are present near the region just below the $DD$ threshold down to ${a_t E_{\sf lhc,\pi\pi}\approx 0.65}$.

It is observed that parameters containing only $D$-wave contributions for the $DD$ and $DD^*$ channels are needed to reproduce the lattice levels from the quantization condition, even though the corresponding $t$-matrix elements are smaller than the $D^*D^*$ $S$-wave across all variations.
The small magnitude of the $D$-waves is also manifest in the possibility of interchangeably fixing them to zero in different parametrizations without significantly different results.
The $DD^*\SLJc{3}{D}{3}$ wave is included in all variations to account for a high-lying energy level but has small magnitude, similar to values extracted from the $T_1^+$ and $A_2^+$ irreps in Sec.~\ref{sec:1p} (see Fig.~\ref{fig:coupled_3p}).

The left panel of Fig.~\ref{fig:coupled_2p} shows the example parametrization as bands, and all the parametrization variations from Table~\ref{tab:par_var_2pcoupled} as curves. It can be seen that the ${D^*D^* \SLJc{5}{S}{2} \to D^*D^* \SLJc{5}{S}{2}}$ amplitude is the dominant one in relation to the $D \to S$-waves also depicted.
The remaining amplitudes containing $D$-wave contributions not shown in this figure are consistent with zero.
No signal of potential $J^P=2^+$ resonances or bound states is observed in the energy region considered.

The right panel of Fig.~\ref{fig:coupled_2p} shows the phases and inelasticities from the generalized Stapp parametrization for the partial-wave amplitudes considered.
The $DD$ and $DD^*$ phase shifts are consistent with zero throughout the energy region considered, and the $D^*D^*$, $S$-wave turns on to a small though non-zero value at the associated threshold.
The sign of the $D^*D^*$, $S$-wave phase shift indicates a repulsive interaction on this channel.
The inelasticity $\cos 2 \bar \epsilon_{ij}$ between the $D$-waves is consistent with $1$.
On the other hand, the inelasticities between the $D$-waves and the $D^*D^*$, $S$-wave are small but significantly differ from unity.

\section{\label{sec:interpret} Interpretation}

The results from the previous section show predominantly weak and repulsive interactions involving $D$ and $D^*$ in $I=1$ and $J^P=0^+,1^+,2^+$, with small coupled-channel effects.
A particular $S$-wave amplitude dominates each of the $J^P=1^+,2^+$ channels below the $DD\pi$ threshold, namely $DD^*\SLJc{3}{S}{1}$ and $D^*D^*\SLJc{5}{S}{2}$.
Furthermore, a repulsive $DD\SLJc{1}{S}{0}$ interaction along with a weakly attractive $D^*D^*\SLJc{1}{S}{0}$ interaction in the coupled channel region are observed in $J^P=0^+$.
There is no signal of resonance, bound-state or virtual bound-state poles in the channels and energy region examined, which is consistent with the $D^0 D^0 \pi^+$ experimental data analysis leading to a preference for the $T_{cc}^+(3875)$ state to be in isospin-$0$~\cite{LHCb:2021auc}.

These results are also consistent with the typical expectations for doubly-charmed scattering in $I=1$ from other theoretical approaches, in particular for the $J^P=1^+$ case where a $T_{cc}^+$-like state is present in the $I=0$, \eg Ref.~\cite{Yang:2019itm}.
Additionally, the similarity between the $S$-wave interactions in ${DD^*, J^P=1^+}$ and ${D^*D^*, J^P=2^+}$ are in agreement with frameworks based on heavy quark spin symmetry~\cite{Albaladejo:2021vln}.
The attractive interaction observed in $D^*D^*$, $J^P=0^+$ is not strong enough to form states, and is compatible with results from one-pion exchange potentials~\cite{Thomas:2008ja}.

There is a qualitative level of consistency between the $DD, J^P=0^+$ and $DD^*, J^P=1^+$ scattering lengths from this work, as quoted in Sections~\ref{sec:0pelastic} and \ref{sec:1p}, and previous studies in the elastic region using different lattice setups~\cite{Ikeda:2013vwa,Chen:2022vpo,Meng:2024kkp,Shi:2025ogt}.
In particular, Ref.~\cite{Meng:2024kkp} studied the two-body isovector $DD^*$, $J^P=1^+$ scattering amplitude by solving Lippmann-Schwinger equations in a non-relativistic effective theory approach, where left-hand cut effects were explicitly included via an one-pion exchange potential.
A naive comparison to the scattering length obtained there suggests that accounting for the left-hand cut does not change the overall weakly-repulsive character of this channel.
Care should be taken as these calculations employ different lattice setups, and thus differences such as quark masses and discretization effects can obscure the comparison.
Nevertheless, a quantitative difference can be expected not only from the use of different lattice setups, but also whether the finite-volume method employed accounts for one-pion exchange effects.
This prompts further studies of the doubly charmed sector using extensions of the Lüscher formalism accounting for left-hand singularities~\cite{Raposo:2023oru,Raposo:2025dkb,Hansen:2024ffk}.

It is also useful to consider the results above in light of SU$(3)_f$ symmetry, where the light and strange quark masses are taken to be equal. 
Following this perspective, the doubly-charmed states fall into the $\bm 3$ and $\bar{\bm 6}$ multiplets of SU$(3)_f$.
These can be decomposed into various $D^{(*)}D^{(*)}$, $D^{(*)} D^{(*)}_s$ and $D^{(*)}_s D^{(*)}_s$ states with definite isospin as follows:
\begin{align}
\begin{split}
    D^{(*)}D^{(*)} (I=0)           \quad : \quad & \bm 3 \,,  \\
    D^{(*)}D^{(*)} (I=1)           \quad : \quad & \bar{\bm 6} \,,  \\
    D^{(*)}_s D^{(*)}_s (I=0)  \quad : \quad & \bar{\bm 6} \,,  \\
    D^{(*)} D^{(*)}_s (I=1/2)  \quad : \quad & \bm 3 + \bar{\bm 6} \,.
\end{split}
\end{align}
In particular, the $D^{(*)}D^{(*)}, I=1$ and $D^{(*)}_s D^{(*)}_s, I=0$ channels are exclusively present in the $\bar{\bm 6}$ irrep.
Supposing that interactions do not change qualitatively away from the SU$(3)_f$ point, then the results of this work would also suggest that $D^{(*)}_s D^{(*)}_s, I=0$ is weakly repulsive.
Under the same suppositions, it is also interesting to note that $D^{(*)} D^{(*)}_s$ in $I=1/2$ appears in the decomposition of both $\bm 3$ and $\bar{\bm 6}$ multiplets, and thus could present a combination of $I=0$ and $I=1$ physics.
In particular, due to the observation of the isoscalar $T_{cc}^+$ state, this could suggest a potential strange partner in ${D^* D_s - D D_s^*, I=1/2}$ (see \eg Ref.~\cite{Karliner:2021wju}).
A lattice calculation on this channel was recently presented in Ref.~\cite{Shrimal:2025ues}.

For completeness, we briefly consider the negative-parity counterpart of the channels above in Appendix~\ref{apx:negparity}.
In this case, only $DD^*$ and $D^*D^*$ are allowed by symmetry, with dominant $P$-wave contributions.
In Fig.~\ref{fig:fvspectraNegP}, the absence of energy shifts, and phases consistent with zero imply no significant interactions in this sector.

\section{\label{sec:conclusions} Conclusions}

In this work, we presented the first coupled-channel calculation involving $DD, DD^*$ and $D^*D^*$ scattering in isospin-$1$ and $J^P=0^+,1^+,2^+$ from lattice QCD.
The interactions in $J^P=1^+,2^+$ were determined to be weakly repulsive, and each of those $J^P$ features a dominant \text{$S$-wave} component of similar magnitude in the energy region investigated.
In $J^P=0^+$, in addition to a repulsive $DD$, $S$-wave interaction, a weakly attractive $D^*D^*$, $S$-wave interaction is observed in the coupled-channel region.
No singularities corresponding to bound states or resonances were found in this energy region. For $J^P=1^+$, this is consistent with the absence of isospin-$1$ states and the identification of the $T_{cc}^+(3875)$ state in the analysis of $D^0D^0\pi^+$ experimental data~\cite{LHCb:2021auc} as isospin-$0$.
The investigation of isospin-$1$ channels also complements the picture started in Ref.~\cite{Whyte:2024ihh}, where a virtual bound state and a resonance were observed in isospin-$0$ using the same calculation setup.
The extraction of the amplitudes in this work will also be an ingredient for future three-body and left-hand cut analyses of $DD\pi$ and $DD^*$ scattering.
On a similar note, the lattice investigation of analogous systems containing $D_s^{(*)}$ mesons can provide further insight into how the physics of the doubly-charmed sector emerges from QCD.

\begin{acknowledgments}
We thank our colleagues within the Hadron Spectrum Collaboration (\url{www.hadspec.org}) for useful discussions.
NPL \& DJW acknowledge support from a Royal Society University Research Fellowship. NPL, DJW \& CET acknowledge support from the U.K. Science and Technology Facilities Council (STFC) [grant numbers ST/T000694/1, ST/X000664/1].

The software codes
{\tt Chroma}~\cite{Edwards:2004sx}, {\tt QUDA}~\cite{Clark:2009wm,Babich:2010mu}, {\tt QUDA-MG}~\cite{Clark:SC2016}, {\tt QPhiX}~\cite{ISC13Phi}, {\tt MG\_PROTO}~\cite{MGProtoDownload}, {\tt QOPQDP}~\cite{Osborn:2010mb,Babich:2010qb}, and {\tt Redstar}~\cite{Chen:2023zyy} were used. 
Some software codes used in this project were developed with support from the U.S.\ Department of Energy, Office of Science, Office of Advanced Scientific Computing Research and Office of Nuclear Physics, Scientific Discovery through Advanced Computing (SciDAC) program; also acknowledged is support from the Exascale Computing Project (17-SC-20-SC), a collaborative effort of the U.S.\ Department of Energy Office of Science and the National Nuclear Security Administration.

This work used the Cambridge Service for Data Driven Discovery (CSD3), part of which is operated by the University of Cambridge Research Computing Service (www.csd3.cam.ac.uk) on behalf of the STFC DiRAC HPC Facility (www.dirac.ac.uk). The DiRAC component of CSD3 was funded by BEIS capital funding via STFC capital grants ST/P002307/1 and ST/R002452/1 and STFC operations grant ST/R00689X/1. Other components were provided by Dell EMC and Intel using Tier-2 funding from the Engineering and Physical Sciences Research Council (capital grant EP/P020259/1). This work also used the earlier DiRAC Data Analytic system at the University of Cambridge. This equipment was funded by BIS National E-infrastructure capital grant (ST/K001590/1), STFC capital grants ST/H008861/1 and ST/H00887X/1, and STFC DiRAC Operations grant ST/K00333X/1. DiRAC is part of the National E-Infrastructure.
This work also used clusters at Jefferson Laboratory under the USQCD Initiative and the LQCD ARRA project.

Propagators and gauge configurations used in this project were generated using DiRAC facilities, at Jefferson Lab, and on the Wilkes GPU cluster at the University of Cambridge High Performance Computing Service, provided by Dell Inc., NVIDIA and Mellanox, and part funded by STFC with industrial sponsorship from Rolls Royce and Mitsubishi Heavy Industries. Also used was an award of computer time provided by the U.S.\ Department of Energy INCITE program and supported in part under an ALCC award, and resources at: the Oak Ridge Leadership Computing Facility, which is a DOE Office of Science User Facility supported under Contract DE-AC05-00OR22725; the National Energy Research Scientific Computing Center (NERSC), a U.S.\ Department of Energy Office of Science User Facility located at Lawrence Berkeley National Laboratory, operated under Contract No. DE-AC02-05CH11231; the Texas Advanced Computing Center (TACC) at The University of Texas at Austin; the Extreme Science and Engineering Discovery Environment (XSEDE), which is supported by National Science Foundation Grant No. ACI-1548562; and part of the Blue Waters sustained-petascale computing project, which is supported by the National Science Foundation (awards OCI-0725070 and ACI-1238993) and the state of Illinois. Blue Waters is a joint effort of the University of Illinois at Urbana-Champaign and its National Center for Supercomputing Applications.
\end{acknowledgments}

\section*{Data Availability}

Reasonable requests for data, such as energy levels and correlations, can be directed to the authors and will be considered in accordance with the Hadron Spectrum Collaboration's policy on sharing data.

\onecolumngrid
\clearpage
\appendix

\section{\label{apx:negparity} Negative parity sector}

Here, the isospin-$1$ doubly charmed results from the main text are complemented with the negative parity sector ($P=-$).
In this case, $DD$ is not allowed due to Bose symmetry, while $P$ waves appear in $DD^*$ and $D^*D^*$ as contributions to the $[000]A_1^-, [000]T_1^-$ and $[000]E^-$ irreps.
Table~\ref{tab:negparity} further details the allowed partial wave contributions to these irreps, neglecting $J>4$.

\begin{figure*}[!h]
    \centering
    \includegraphics[width=0.3025\textwidth, trim={2mm  2mm 2mm 2mm},clip]{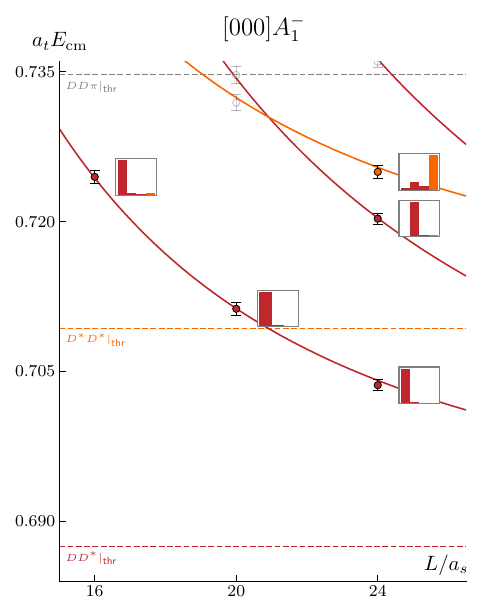}
    \includegraphics[width=0.32\textwidth, trim={2mm  2mm 2mm 2mm},clip]{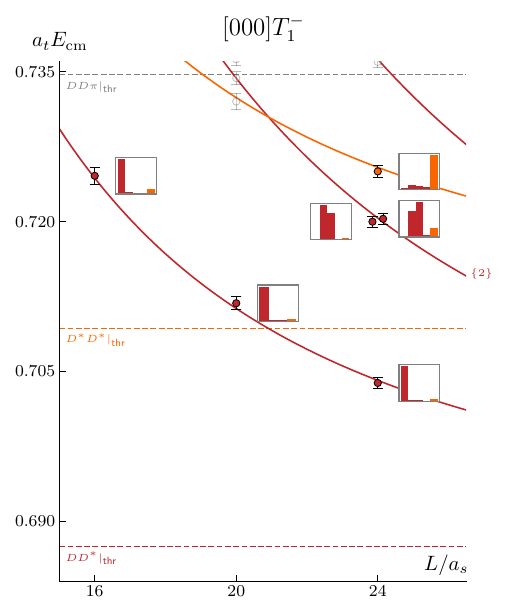}
    \includegraphics[width=0.32\textwidth, trim={2mm  2mm 2mm 2mm},clip]{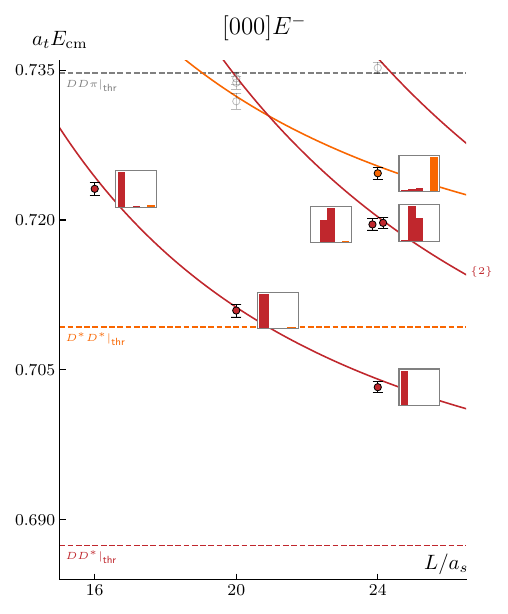}
    
    \includegraphics[width=0.32\textwidth, trim={-0.4mm  2mm 0.5mm 0.5mm},clip]{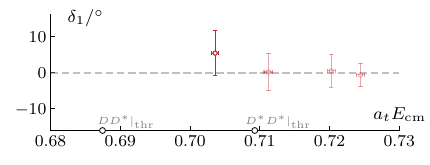}
    \includegraphics[width=0.32\textwidth, trim={3.25mm 2mm 0.5mm 0.5mm},clip]{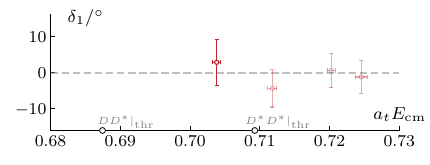}
    \includegraphics[width=0.32\textwidth, trim={3mm  2mm 0.5mm 0.5mm},clip]{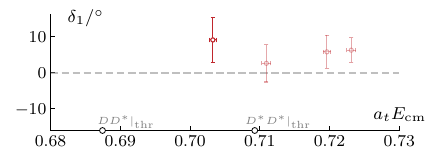}
    \caption{\label{fig:fvspectraNegP}
    Upper: Same as Figs.~\ref{fig:fvspectra1} and \ref{fig:fvspectra2}, but for the rest-frame irreps $[000]A_1^-$, $[000]T_1^-$ and $[000]E^-$.
    Lower: from left to right, phase shifts for the negative-parity $DD^*$, $\SLJ{3}{P}{0}$, $\SLJ{3}{P}{1}$ and $\SLJ{5}{P}{2}$ channels evaluated via Eq.~\ref{eq:luscher} at the lowest $DD^*$-like energies (red) on each of the $A_1^-$, $T_1^-$ and $E^-$ irreps, respectively.
    The translucent phase shift points were evaluated assuming elastic scattering.
    In $T_1^-$ and $E^-$, only one of the values around $a_t \Ecm \approx 0.71$ is shown in the phase-shift panels.
    }
\end{figure*}

The upper panel of Fig.~\ref{fig:fvspectraNegP} shows the finite-volume spectra and the corresponding operator-state overlaps for the $A_1^-, T_1^-$ and $E^-$ irreps, in the same fashion of Fig.~\ref{fig:fvspectra1}.
The lower panel of Fig.~\ref{fig:fvspectraNegP} shows the $P$-wave phase shifts computed at the $DD^*$-like energies (red), ignoring partial waves with $\ell>2$ and assuming the coupling of $DD^*$ and $D^*D^*$ to be negligible.
This can be reasonably justified by the near-threshold suppression of $\ell\geq3$ waves, and by the observation of a small mixing between $DD^*$ and $D^*D^*$ on the operator overlaps in finite volume.
The shifts of the lattice energies to noninteracting energies do not suggest any significant $DD^*$ or $D^*D^*$ interactions in $\SLJ{3}{P}{0,1,2}$.
Furthermore, the observation of $DD^*$ weak interactions is reinforced by the small phase shift values.

\begin{table}[!h]
    \centering
    \setlength{\tabcolsep}{0.5em}{\renewcommand{\arraystretch}{1.5}{%
    \begin{tabular}{c||c|c}
        $[n_x n_y n_z] \Lambda^{(P)}$  & $DD^*$  and $D^*D^*$                             \\\hline\hline
        $[000]A_1^-$                   & $\SLJ{3}{P}{0}$, $\SLJ{3}{F}{4}$, $\SLJ{3}{H}{4}$  \\\hline
        $[000]T_1^-$                   & $\SLJ{3}{P}{1}$, $\SLJ{3}{F}{3}$                   \\\hline
        $[000]E^-$                     & $\SLJ{3}{P}{2}$, $\SLJ{3}{F}{2}$, $\SLJ{3}{F}{4}$, $\SLJ{3}{H}{4}$ 
    \end{tabular}
    }}%
    \caption{
        \label{tab:negparity}
        Contribution of $I=1$, $D^{(*)}D^{(*)}$ partial waves to irreps with negative parity ($P=-$), up to $J=4$.
        The last column follows the spectroscopic notation $\SLJ{2s+1}{\ell}{J}$.
    }
\end{table}

\clearpage
\section{\label{apx:polepush} Procedure for Dealing with Spurious Bound States}

In Sec.~\ref{sec:scattering}, some amplitude parametrizations can feature bound-state poles when extrapolated too far from the lattice energy levels. 
These poles can appear even though no corresponding energy levels are present in the lattice calculation, and thus are seen as artifacts from the specific parametrizations.
Here, we detail the procedure for extracting scattering amplitude parametrizations while ``smoothly pushing'' these singularities away from the energy region considered.

The procedure consists of adding an extra term to the usual spectrum chi-squared, the latter given in Ref.~\cite{Wilson:2014cna}.
For each bound-state pole found in the $t$-matrix at energy $E_\mathrm{p}$, \ie on the first Riemann sheet ($\mathrm{Im}\,k >0$) at purely real values of $\sqrt{s} = E_\mathrm{p}$, the added term has the form
\begin{equation}
    \chisqadd (E_\mathrm{p}) = \alpha \left[ \frac{f(E_\mathrm{p})}{1+ e^{-C(E_\mathrm{p})}} \right]^2
    \label{eq:chisqadd}
\end{equation}
with the functions
\begin{equation}
    f(E_\mathrm{p}) = \frac{E_\mathrm{p}^2 - E_l^2}{E_h^2 - E_l^2}, \qquad C(E_\mathrm{p}) = \frac{E_\mathrm{p} - (E_h + E_l)/2}{E_h - E_l} \,,
\end{equation}
and the constant $\alpha > 0$.
The function $f$ penalizes the occurrence of bound-state poles on the energy interval $[E_l, E_h]$.
As discussed in Sec.~\ref{sec:scattering}, this is justified given that the lattice calculation does not yield any finite-volume energies below the lowest kinematic threshold.

\begin{figure*}[!b]
    \centering
    \includegraphics[width=0.99\linewidth]{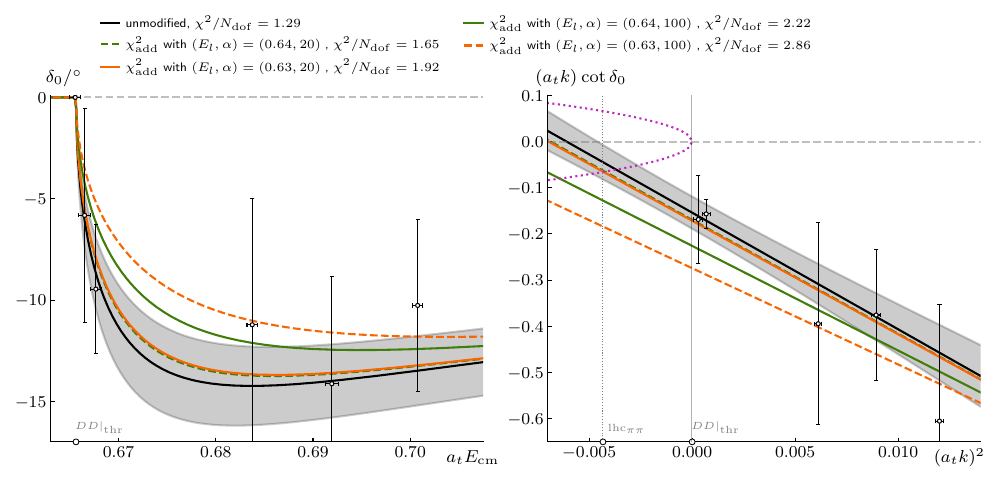}
    \caption{\label{fig:polexc_demo}
    Similar to Fig.~\ref{fig:elastic0pbands}, but comparing the phase shifts (left) and $k \cot \delta_0$ (right) of the effective range parametrization applied to $DD$, $I=1$ elastic scattering in $J^P=0^+$, obtained by the use of $\chisqadd$ at various choices of $\alpha$ and $E_l$.
    For clarity, only the statistical band (gray) of the unmodified parametrization is shown.
    The phase-shift points are derived from the lattice energies using Eq.~\eqref{eq:luscher}.
    The dotted purple lines are exactly as in Fig.~\ref{fig:elastic0pbands}.
    }
\end{figure*}

Specifically in this work, we choose $E_h$ to be the lowest two-particle threshold and $E_l$ to be the first left-hand branch point appearing below $E_h$, the latter associated with the one-particle exchange between the same two particles.
In practice, $E_l$ is tuned to lower values to ensure that the statistical uncertainties of the spurious poles are taken into account on each particular parametrization.
When this is employed, the fit is a result of the minimization of the total chi-squared.
The value of $\alpha$ is somewhat arbitrary, but one could expect that it should be at least of the order of the number of degrees of freedom to cause a visible effect.
In this work, we find that the fixed value $\alpha=100$ works well with the choice of $[E_l, E_h]$ described above on all parametrizations where it is used.
We note that resulting parametrizations are consistent with the one where $\chisqadd$ was not used in the energy region of interest, and that the data is still reasonably described.
Whenever the use of $\chisqadd$ is indicated in this work, the $\chi^2$ value reported will implicitly include $\chisqadd$, \ie $\chi^2 \to \chi^2 + \chisqadd$.

The effect of this procedure is illustrated in Fig.~\ref{fig:polexc_demo} in a case where a spurious bound state appears near the left-hand branch point in $J^P=0^+$, $DD$ elastic scattering when employing the effective range parametrization (see Sec.~\ref{sec:0pelastic}).
The use of the $\chisqadd$ term in the fit procedure results in a parametrization with a spurious pole constrained to be further away from the energy region where the lattice data is present.
This is indicated by the intersections of the $k \cot \delta_0$ curves with the dotted line at lower values of $(a_t k)^2$.
For comparison, we show the combination of two choices of $E_l$ and $\alpha$ and their effect on the central value of $k \cot \delta_0$.
The resulting parametrizations have progressively higher $\chi^2/ N_\mathrm{dof}$ as $\alpha$ is increased and $E_l$ is decreased, as expected from the addition of $\chisqadd$ to the unmodified $\chi^2$.
In particular, for the choices $(E_l,\alpha)=(0.63,100)$ and $(0.63,20)$ the results start to significantly differ from the unmodified parametrization.
At the same time, the former does not describe the lattice levels near threshold as well as the other choices.
In the end, the value $(E_l,\alpha)= (0.64, 100)$ is used for $J^P=0^+$, $DD$ elastic scattering when necessary, as reported in Table~\ref{tab:par_var_0pelastic}.
A similar behavior is observed for the parametrizations in all other cases in this work where the above procedure is used.
The value $\alpha=100$ is kept fixed throughout, and $E_l$ is as specified in Tables~\ref{tab:par_var_0pcoupled} and \ref{tab:par_var_2pcoupled}.

\onecolumngrid
\section{Tables of Operators}
\label{apx:operatortables}

In this appendix, we list all the two-hadron operators $D^{(*)}[\vec{p_1}] D^{(*)}[\vec{p_2}]$ used in this work, in their respective cubic symmetry irreps and lattice spatial extension.
The spatial momenta are given in units of $2\pi/L$ and the number in curly brackets indicates the multiplicity coming from different combinations of single-hadron irreps, if this is greater than one.
The row ordering from top to bottom corresponds first to the opening of the kinematic threshold of the associated hadron-hadron channel, and then by associated noninteracting energy, thus matching the histogram bar ordering in Figs.~\ref{fig:fvspectra1} and \ref{fig:fvspectra2}.
Correlation functions involving all operators were computed and analyzed in the GEVP, but only energy levels corresponding to the operators in bold in Tables~\ref{tab:operatortable0p},~\ref{tab:operatortable1p},\ref{tab:operatortable2p} were used in the scattering amplitude determinations.
The energy levels associated to the operators in normal font were not used in this work.

The operators for the negative-parity irreps are listed in Tables~\ref{tab:operatortable0m1m} and \ref{tab:operatortable2m}.
The ones signaled with ``$(*)$'' correspond to the colored energy levels in the upper panels of Fig.~\ref{fig:fvspectraNegP}.

\begin{table}[!h]
    \begin{minipage}{.99\linewidth}
        \footnotesize
        \setlength{\tabcolsep}{0.3em}{\renewcommand{\arraystretch}{1.5}{
        \begin{tabular}{c|c|c}
            \multicolumn{3}{c}{$[000]A_1^+$}    \\\hline
            $L/a_s=16$                      & $L/a_s=20$                        & $L/a_s=24$    \\\hline
            \bm{$D[000] D[000]$}            & \bm{$D[000] D[000]$}              & \bm{$D[000] D[000]$} \\
            \bm{$D[100] D[100]$}            & \bm{$D[100] D[100]$}              & \bm{$D[100] D[100]$} \\
            \bm{$D[110] D[110]$}            & \bm{$D[110] D[110]$}              & \bm{$D[110] D[110]$} \\
            \bm{$D[111] D[111]$}            & \bm{$D[111] D[111]$}              & \bm{$D[111] D[111]$} \\
            $D[200] D[200]$                 & --                                & $D[200] D[200]$ \\
            \bm{$D^*[000] D^*[000]$}        & \bm{$D^*[000] D^*[000]$}          & \bm{$D^*[000] D^*[000]$} \\
            $\{2\} D^*[100] D^*[100]$       & $\{2\} D^*[100] D^*[100]$         & \bm{$\{2\} D^*[100] D^*[100]$} \\
        \end{tabular}
        }}
    \end{minipage}
    \caption{\label{tab:operatortable0p} Operators in the $[000]A_1^+$ irrep, used to extract $J^P=0^+,4^+$ amplitudes.}
\end{table}

\bigskip
\onecolumngrid

\begin{table*}[h!]
    \centering
    \begin{minipage}{.49\linewidth}
        \footnotesize
        \setlength{\tabcolsep}{0.3em}{\renewcommand{\arraystretch}{1.5}{
        \begin{tabular}{c|c|c}
            \multicolumn{3}{c}{$[000]T_1^+$}    \\\hline
            $L/a_s=16$                      & $L/a_s=20$                        & $L/a_s=24$ \\\hline
            \bm{$D[000] D^*[000]$}          & \bm{$D[000] D^*[000]$}            & \bm{$D[000] D^*[000]$}        \\
            \bm{$\{2\} D[100] D^*[100]$}    & \bm{$\{2\} D[100] D^*[100]$}      & \bm{$\{2\} D[100] D^*[100]$}   \\
            $\{3\}D[110] D^*[110]$          & $\{3\} D[110] D^*[110]$           & \bm{$\{3\}D[110] D^*[110]$}    \\
            $\{2\}D[111] D^*[111]$          & -                                 & $\{2\}D[111] D^*[111]$    \\
            $D^*[100] D^*[100]$             & $D^*[100] D^*[100]$               & \bm{$D^*[100] D^*[100]$} \\
            & &  \\
            & &  \\
        \end{tabular}
        }}
    \end{minipage}
    \begin{minipage}{.2\linewidth}
        \centering
        \footnotesize
        \setlength{\tabcolsep}{0.3em}{\renewcommand{\arraystretch}{1.5}{
        \begin{tabular}{c}
            \multicolumn{1}{c}{$[000]A_2^+$}    \\\hline
            $L/a_s=24$ \\ \hline
            $D[100] D^*[100]$   \\
            \bm{$D[111] D^*[111]$}   \\
            $D^*[100] D^*[100]$ \\
            \\
            \\
            \\
            \\
        \end{tabular}
        }}
    \end{minipage}
    \caption{\label{tab:operatortable1p} Operators in the $[000]T_1^+$ and $[000]A_2^+$ irreps, used to extract $J^P=1^+,3^+$ amplitudes.}
\end{table*}

\begin{table*}[h!]
    \centering
    \begin{minipage}{.49\linewidth}
        \footnotesize
        \setlength{\tabcolsep}{0.3em}{\renewcommand{\arraystretch}{1.5}{
        \begin{tabular}{c|c|c}
            \multicolumn{3}{c}{$[000]E^+$}    \\\hline
            $L/a_s=16$                  & $L/a_s=20$                    & $L/a_s=24$                \\\hline
            \bm{$D[100] D[100]$}        & \bm{$D[100] D[100]$}          & \bm{$D[100] D[100]$}           \\
            $D[110] D[110]$             & \bm{$D[110] D[110]$}          & \bm{$D[110] D[110]$}           \\
            $D[200] D[200]$             & -                             & $D[200] D[200]$           \\
            $D[110] D^*[110]$           & $D[110] D^*[110]$             & \bm{$D[110] D^*[110]$}         \\
            $D[111] D^*[111]$           & -                             & $D[111] D^*[111]$         \\
            \bm{$D^*[000] D^*[000]$}    & \bm{$D^*[000] D^*[000]$}      & \bm{$D^*[000] D^*[000]$}       \\
            $\{3\} D^*[100] D^*[100]$   & $\{3\} D^*[100] D^*[100]$     & \bm{$\{3\} D^*[100] D^*[100]$} \\
        \end{tabular}
        }}
    \end{minipage}\quad
    \begin{minipage}{.49\linewidth}
        \footnotesize
        \setlength{\tabcolsep}{0.3em}{\renewcommand{\arraystretch}{1.5}{
        \begin{tabular}{c|c|c}
            \multicolumn{3}{c}{$[000]T_2^+$}    \\\hline
            $L/a_s=16$                  & $L/a_s=20$                & $L/a_s=24$                \\\hline
            $D[110] D[110]$             & \bm{$D[110] D[110]$}      & \bm{$D[110] D[110]$}           \\
            $D[111] D[111]$             & $D[111] D[111]$           & \bm{$D[111] D[111]$}           \\
            \bm{$D[100] D^*[100]$}      & \bm{$D[100] D^*[100]$}    & \bm{$D[100] D^*[100]$}         \\
            $\{2\} D[110] D^*[110]$     & $\{2\} D[110] D^*[110]$   & \bm{$\{2\} D[110] D^*[110]$}   \\
            $D[111] D^*[111]$           & -                         & $D[111] D^*[111]$         \\
            \bm{$D^*[000] D^*[000]$}    & \bm{$D^*[000] D^*[000]$}  & \bm{$D^*[000] D^*[000]$}       \\
            $\{2\} D^*[100] D^*[100]$   & $\{2\} D^*[100] D^*[100]$ & \bm{$\{2\} D^*[100] D^*[100]$} \\
        \end{tabular}
        }}
    \end{minipage}
    \caption{\label{tab:operatortable2p} Operators in the $[000]E^+$ and $[000]T_2^+$ irreps, used to extract $J^P=2^+$ amplitudes.}
\end{table*}

\begin{table}[!h]
    \begin{minipage}{.49\linewidth}
        \footnotesize
        \setlength{\tabcolsep}{0.3em}{\renewcommand{\arraystretch}{1.5}{
        \begin{tabular}{c|c|c}
            \multicolumn{3}{c}{$[000]A_1^-$}    \\\hline
            $L/a_s=16$                      & $L/a_s=20$                        & $L/a_s=24$    \\\hline
            $D[100] D^*[100]\ (*)$           & $D[100] D^*[100]\ (*)$             & $D[100] D^*[100]\ (*)$ \\
            $D[110] D^*[110]$               & $D[110] D^*[110]$                 & $D[110] D^*[110]\ (*)$ \\
            $D[111] D^*[111]$               & -                                 & $D[111] D^*[111]$ \\
            $D^*[100] D^*[100]$             & $D^*[100] D^*[100]$               & $D^*[100] D^*[100]\ (*)$ \\
        \end{tabular}
        }}
    \end{minipage}\quad
    \begin{minipage}{.49\linewidth}
        \footnotesize
        \setlength{\tabcolsep}{0.3em}{\renewcommand{\arraystretch}{1.5}{
        \begin{tabular}{c|c|c}
            \multicolumn{3}{c}{$[000]T_1^-$}    \\\hline
            $L/a_s=16$                      & $L/a_s=20$                        & $L/a_s=24$    \\\hline
            $D[100] D^*[100]\ (*)$           & $D[100] D^*[100]\ (*)$             & $D[100] D^*[100]\ (*)$ \\
            $ \{2\}D[110] D^*[110]$         & $ \{2\}D[110] D^*[110]$           & $\{2\} D[110] D^*[110]\ (*)$ \\
            $D[111] D^*[111]$               & -                                 & $D[111] D^*[111]$ \\
            $ \{2\}D^*[100] D^*[100]$       & $D^*[100] D^*[100]$               & $D^*[100] D^*[100]\ (*)$ \\
        \end{tabular}
        }}
    \end{minipage}
    \caption{\label{tab:operatortable0m1m}
    Operators in the $[000]A_1^-$ and $[000]T_1^-$ irreps. 
    The operators signaled with ``$(*)$'' correspond to the energy levels highlighted in the upper panels of Fig.~\ref{fig:fvspectraNegP}.
    }
\end{table}

\begin{table}[!t]
    \begin{minipage}{.99\linewidth}
        \footnotesize
        \setlength{\tabcolsep}{0.3em}{\renewcommand{\arraystretch}{1.5}{
        \begin{tabular}{c|c|c}
            \multicolumn{3}{c}{$[000]E^-$}    \\\hline
            $L/a_s=16$                      & $L/a_s=20$                        & $L/a_s=24$    \\\hline
            $D[100] D^*[100]\ (*)$           & $D[100] D^*[100]\ (*)$             & $D[100] D^*[100]\ (*)$ \\
            $\{2\}D[110] D^*[110]$          & $\{2\}D[110] D^*[110]$            & $\{2\} D[110] D^*[110]\ (*)$ \\
            $D[111] D^*[111]$               & -                                 & $D[111] D^*[111]$ \\
            $D^*[100] D^*[100]$             & $D^*[100] D^*[100]$               & $D^*[100] D^*[100]\ (*)$ \\
        \end{tabular}
        }}
    \end{minipage}
    \caption{\label{tab:operatortable2m}
    Same as Table~\ref{tab:operatortable0m1m}, but for the $[000]E^-$ irrep.}
\end{table}

\vspace{40cm}

\clearpage
\section{Tables of Parametrization Variations}
\label{apx:partables}

Tables~\ref{tab:par_var_0pelastic}-\ref{tab:par_var_2pcoupled} detail the scattering amplitude parametrizations in isospin-$1$, as presented in Sec.~\ref{sec:scattering}.
Each row corresponds to a certain choice of K-matrix parameters of the expansion Eq.~\eqref{eq:k_poly} or an effective range expansion.
In all tables except for Table~\ref{tab:par_var_1pelastic}, the example parametrizations quoted in Eqs.~\eqref{eq:0pelasticpars}-\eqref{eq:2pcoupledpars} are shown highlighted.
For each spectrum chi-squared minimization performed, the corresponding $\chi^2/ N_\mathrm{dof}$ is shown in the right-most columns.
When the spurious bound-state treatment from Appendix~\ref{apx:polepush} is employed during this minimization, the quoted $\chi^2$ includes the extra $\chisqadd$ term, and corresponds to a separate fit.
In every case $\chisqadd$ is used, the parameter $\alpha=100$ from Eq.~\eqref{eq:chisqadd} is fixed.

Note that for all parametrizations listed, we ensure that unphysical poles are not present in the region of the complex plane constrained by the energy levels.
The only exceptions are the last two parametrizations on Table~\ref{tab:par_var_0pelastic}, which are left as examples of where this problem can happen.

\begin{table}[!h]
    \centering
    \setlength{\tabcolsep}{0.3em}{\renewcommand{\arraystretch}{2}{%
    \footnotesize
    \caption{\label{tab:par_var_0pelastic}
    Variations of effective-range and K-matrix scattering parametrizations for $I=1$, elastic $DD$ scattering in $J^P=0^+$, extracted from fits to the lattice energies below the $D^*D^*$ threshold, as described in Sec.~\eqref{sec:0pelastic}. 
    In the cases where $\chisqadd$ was employed, the value of the parameter $a_t E_l=0.64$ is used (see Eq.~\eqref{eq:chisqadd}).
    For simplicity, we use the shorthand $\gamma^{(n)} \equiv \gamma^{(n)}_{ DD\SLJc{1}{S}{0} \to DD\SLJc{1}{S}{0} }$.
    The parametrizations signaled with a ``$\ddagger$'' have high $\chi^2/ N_\mathrm{dof}$ values and are only given for completeness, but are not considered in any further analysis or interpretation.
    The ones with a ``$\dagger$'' are not considered reasonable descriptions of the data due to the presence of a spurious bound-state pole near the region constrained by the data.
    The parametrizations denoted by ``$\P$'' have nearby physical-sheet poles with nonzero imaginary parts and are also discarded.
    The remaining results are reasonable parametrizations and correspond to the curves in Fig.~\ref{fig:elastic0pbands}.
    The parametrization denoted by ``$*$'' is an example parametrization, represented by the bands in Fig.~\ref{fig:elastic0pbands} and quoted in Eq.~\eqref{eq:0pelasticpars}.
    }
    \begin{tabular}{c|c|c}
    \hline
    Parameterization & \multicolumn{2}{c}{$\chi^2/ N_\mathrm{dof}$} \\
    \hline\hline
    & without $\chisqadd$ & with $\chisqadd$ \\\hline

    \multicolumn{3}{c}{K-matrix} \\\hline

    Chew-Mandelstan phase space & & \\\hline

    $K(\hat s) = \gamma^{(0)}$                                          & $\frac{22.91}{(6 - 1)}=4.58 \ ^\ddagger$  & --   \\
    $\bm{K(\hat s) = \gamma^{(0)} + \gamma^{(1)} \hat s}$               & $\bm{\frac{2.81 }{(6 - 2)}=0.70 \ ^*}$    & --   \\
    $K(\hat s) = \gamma^{(0)} + \gamma^{(2)} \hat s^2$                  & $\frac{2.38 }{(6 - 2)}=0.59$   & --   \\
    $K(\hat s) = \gamma^{(0)} + \gamma^{(1)} + \gamma^{(2)} \hat s^2$   & $\frac{1.70 }{(6 - 3)}=0.57$   & --   \\

    \hline
    Simple phase space &  &  \\\hline

    $K(\hat s) = \gamma^{(0)}$                                          & $\frac{22.25}{(6 - 1)}=4.45 \ ^\ddagger$  & --    \\
    $K(\hat s) = \gamma^{(0)} + \gamma^{(1)} \hat s$                    & $\frac{2.44 }{(6 - 2)}=0.61$  & --    \\
    $K(\hat s) = \gamma^{(0)} + \gamma^{(2)} \hat s^2$                  & $\frac{2.76 }{(6 - 2)}=0.69$  & --    \\
    $K(\hat s) = \gamma^{(0)} + \gamma^{(1)} + \gamma^{(2)} \hat s^2$   & $\frac{1.74 }{(6 - 3)}=0.58$  & --    \\
    \hline\hline

    \multicolumn{3}{c}{Effective Range} \\\hline

    $k \cot \delta(k) = 1/a $                                           & $\frac{23.27}{(6 - 1)}=4.65$              & --    \\
    $k \cot \delta(k) = \frac{1/a}{1-k^2/\kappa^2} $                    & $\frac{2.60}{(6 - 2)}=0.65$               & --    \\
    $k \cot \delta(k) = 1/a +rk^2 /2 $                                  & $\frac{5.17}{(6 - 2)}=1.29 \ ^\dagger$    & $\frac{8.87}{(6 - 2)}=2.22$    \\
    $k \cot \delta(k) = \frac{1/a + rk^2/2}{1-k^2/\kappa^2}$            & $\frac{1.80}{(6 - 3)}=0.60 \ ^\dagger$    & $\frac{1.86}{(6 - 3)}=0.62$    \\
    $k \cot \delta(k) = 1/a + P_2 k^4 $                                 & $\frac{2.65}{(6 - 2)}=\mathit{0.66} \ ^\P$         & --    \\
    $k \cot \delta(k) = 1/a +rk^2 /2 + P_2 k^4 $                        & $\frac{1.68}{(6 - 3)}=\mathit{0.56} \ ^\P$         & --    \\
\end{tabular}

    }}%
\end{table}

\begin{table*}[!ht]
    \centering
    \setlength{\tabcolsep}{0.35em}{\renewcommand{\arraystretch}{1.65}{%
    \footnotesize
    \caption{\label{tab:par_var_0pcoupled}
    Variations of the $K$-matrix parametrizations for coupled-channel $DD - D^*D^*$, $I=1$ scattering in $J^P=0^+$ extracted from fits to lattice energies below the $DD\pi$ threshold and whose $\chi^2/ N_\mathrm{dof}$ is given in the last column (see Sec.~\ref{sec:0pcoupled}).
    Some variations also parametrize the $J^P=4^+$ contributions in $[000]A_1^+$.
    The entries of the table display the values of $n$ for which $\gamma_{ij}^{(n)} \neq 0$ in Eq.~\eqref{eq:k_poly}.
    The ``$-$'' or the absence of an amplitude means that the corresponding K-matrix parameters are exactly zero.
    In the cases where $\chisqadd$ was employed, the value of the parameter $E_l=0.63$ is used (see Eq.~\eqref{eq:chisqadd}).
    The parametrizations signaled with a ``$\dagger$'' are not considered reasonable descriptions of the data due to the presence of a spurious bound-state pole above the left-hand cut just below the $DD$ threshold.
    In the ones with $\chi^2/ N_\mathrm{dof}$ set to ``$-$''  the spurious bound-state procedure was not employed.
    The row in bold and denoted by ``$*$'' is an example parametrization, quoted in Eq.~\eqref{eq:0pcoupledpars} and represented in Fig.~\ref{fig:coupled_0p} together with all other reasonable parametrizations.
    }
    
\begin{tabular}{c|c|c|c|c|c||c|c}
    \hline
    $DD$\SLJc{1}{S}{0} & $DD$\SLJc{1}{S}{0} & $DD$\SLJc{1}{S}{0} & $D^*D^*$\SLJc{1}{S}{0} & $D^*D^*$\SLJc{1}{S}{0} & $D^*D^*$\SLJc{5}{D}{4} & \multicolumn{2}{c}{\multirow{2}{4em}{$\chi^2/ N_\mathrm{dof}$}}  \\
    $\to DD$\SLJc{1}{S}{0} & $\to D^*D^*$\SLJc{1}{S}{0} & $\to D^*D^*$\SLJc{5}{D}{0} & $\to D^*D^*$\SLJc{1}{S}{0}  & $\to D^*D^*$\SLJc{5}{D}{0} & $\to D^*D^*$\SLJc{5}{D}{4} & \multicolumn{2}{c}{} \\
    
    \hline\hline
    \multicolumn{6}{c||}{Chew-Mandelstam phase space} & without $\chisqadd$ & with $\chisqadd$ \\\hline
    $0$   & $0$ & $0$ & $0$ & -- & --   & $\frac{12.21}{ (14 - 4)} = 1.22\ ^\dagger$ & $\frac{15.25}{ (14 - 4)} = 1.52$ \\
    $0,1$ & --  & $0$ & $0$ & -- & --   & $\frac{2.94 }{ (14 - 4)} = 0.29\ ^\dagger$ & $\frac{5.29 }{ (14 - 4)} = 0.53$  \\
    $0,1$ & -- & $0$ & $1$ & -- & --  &   $\frac{3.58 }{(14 - 4) } = 0.36\ ^\dagger$ & $\frac{5.33 }{(14 - 4) } = 0.53$ \\
    $0$   & $0$ & $0$ & $0$ & $0$ & --  & $\frac{12.13}{ (14 - 5)} = 1.35\ ^\dagger$ & $\frac{15.12}{ (14 - 5)} = 1.68$ \\
    $0,1$   & -- & -- & $0$ & -- & $0$   & $\frac{16.79}{ (14 - 4)} = 1.68$ & -- \\
    $0,1$ & $0$ & $0$ & $0$ & -- & --   & $\frac{2.84}{ (14 - 5)} =  0.32$ & -- \\
    $0,1$   & $1$ & -- & $0$ & -- & $0$   & $\frac{2.91}{ (14 - 5)} = 0.32$ & -- \\

    \hline\hline
    \multicolumn{6}{c||}{Simple phase space} & & \\\hline
    $\bm{0}$ & $\bm{0}$ & $\bm{0}$ & $\bm{0}$ & -- & -- &  $\bm{\frac{9.79 }{(14 - 4) } = 0.98 \ ^*}$ & -- \\
    $0,1$ & -- & $0$ & $0$ & -- & --  &                    $\frac{3.35 }{(14 - 4) } = 0.33$  & -- \\
    $0,1$ & -- & $0$ & $1$ & -- & --  &                    $\frac{4.12 }{(14 - 4) } = 0.41$  & -- \\
    $0$ & $0$ & $0$ & $0$ & $0$ & --  &                    $\frac{9.75 }{(14 - 5) } = 1.08$  & -- \\
    $0,1$ & -- & -- & $0$ & -- & $0$   &                    $\frac{15.53}{ (14 - 4)} = 1.55$ & -- \\
    $0,1$ & $0$ & $0$ & $0$ & -- & --  &                   $\frac{3.24 }{(14 - 5) } = 0.36$  & -- \\
    $0,1$   & $1$ & -- & $0$ & -- & $0$   & $\frac{3.35}{ (14 - 5)} = 0.37$ & -- \\
\end{tabular}
    }}%
\end{table*}

\begin{table*}[!ht]
    \centering
    \setlength{\tabcolsep}{0.75em}{\renewcommand{\arraystretch}{2}{%
    \footnotesize
    \caption{\label{tab:par_var_1pelastic}
    Analogous to Table~\ref{tab:par_var_0pelastic}, but for parametrization variations of $DD^*$ scattering in $J^P=1^+$ and $I=1$ (see Sec.~\ref{sec:1p}).
    The ``$\#$'' denotes that only three lowest energies were used, contrarily to the other two parametrizations, as described in the text. 
    }
    \begin{tabular}{c|c|c}
    \hline
    Parameterization & \multicolumn{2}{c}{$\chi^2/ N_\mathrm{dof}$} \\
    \hline\hline

    \multicolumn{3}{c}{Effective Range} \\\hline
    & without $\chisqadd$                            & with $\chisqadd$ \\\hline
    $k \cot \delta(k) = 1/a  \ ^\# $                    & $\frac{1.00}{(3 - 1)}=0.50$               & - \\
    $k \cot \delta(k) = 1/a +rk^2 /2 $                  & $\frac{5.86}{(6 - 2)}=1.46 \ ^\dagger$    & $\frac{16.38}{(6 - 2)}=4.10\ ^\ddagger$ \\
    $k \cot \delta(k) = \frac{1/a}{1-k^2/\kappa^2} $    & $\frac{3.85}{(6 - 2)}=0.96$               & - \\

\end{tabular}

    }}%
    
\end{table*}

\begin{table*}[!ht]
    \centering
    \setlength{\tabcolsep}{0.5em}{\renewcommand{\arraystretch}{1.65}{%
    \footnotesize
    \caption{\label{tab:par_var_1p3pcoupled}
    Analogous to Table~\ref{tab:par_var_0pcoupled}, but for parametrization variations of coupled-channel $DD^* -D^*D^*$ scattering in $J^P=1^+$ and $I=1$ (see Sec.~\ref{sec:1p}).
    The parametrization of the $J^P=3^+$ contribution in $[000]T_1^+$ and $[000]A_2^+$ is also shown.
    The example parametrization (``$*$'') is quoted in Eq.~\eqref{eq:1p3pcoupledpars}, and shown together with all other variations in Fig.~\ref{fig:coupled_1p3p}. 
    }
    
\begin{tabular}{c|c|c|c|c|c||c}
    \hline
    $DD^*$\SLJc{3}{S}{1}     & $DD^*$\SLJc{3}{D}{1}     & $DD^*$\SLJc{3}{D}{1}     & $DD^*$\SLJc{3}{D}{3}     & $DD^*$\SLJc{3}{D}{3}       & $D^*D^*$\SLJc{5}{D}{3}      & \multirow{2}{4em}{$\chi^2/ N_\mathrm{dof}$} \\
    $\to DD^*$\SLJc{3}{S}{1} & $\to DD^*$\SLJc{3}{D}{1} & $\to DD^*$\SLJc{3}{S}{1} & $\to DD^*$\SLJc{3}{D}{3} & $\to D^*D^*$\SLJc{5}{D}{3} & $\to D^*D^*$\SLJc{5}{D}{3} & \\

    \hline\hline
    \multicolumn{6}{c}{Chew-Mandelstam phase space} \\\hline

    $0,1$ & $0$   & --    & $0$ & --  & $0$ & $\frac{12.59}{ (15 - 5) } = 1.26$  \\
    $0,1$ & $0,1$ & --    & $0$ & --  & $0$ & $\frac{9.78 }{ (15 - 6) } = 1.09$  \\
    $0,1$ & $0$   & --    & $0$ & $0$ & --  & $\frac{9.07 }{ (15 - 5) } = 0.91$  \\
    $0,1$ & $0$   & $0,1$ & $0$ & $0$ & --  & $\frac{9.07 }{ (15 - 7) } = 1.13$  \\
    \bm{$0,1$} & \bm{$0$}   & \bm{$1$}   & \bm{$0$} & \bm{$0$} & --  & \bm{$\frac{9.07 }{ (15 - 6) } = 1.01 ^*$}  \\
    $0,1$ & $0$   & $0$   & $0$ & --  & $0$ & $\frac{7.00 }{ (15 - 6) } = 0.78$  \\

    \hline\hline
    \multicolumn{6}{c}{Simple phase space} \\\hline

    $0,1$ & $0$   & --  & $0$   & --  & $0$ & $\frac{12.25}{ (15 - 5) } = 1.23$  \\
    $0,1$ & $0,1$ & --  & $0$   & --  & $0$ & $\frac{9.29 }{ (15 - 6) } = 1.03$  \\
    $0,1$ & $0$   & --  & $0$   & $0$ & --  & $\frac{8.66 }{ (15 - 5) } = 0.87$  \\
    $0,1$ & $0$   & $0,1$ & $0$ & $0$ & --  & $\frac{8.66 }{ (15 - 7) } = 1.08$  \\
    $0,1$ & $0$   & $1$ & $0$   & $0$ & --  & $\frac{8.66 }{ (15 - 6) } = 0.96$  \\
    $0,1$ & $0$   & $0$ & $0$   & --  & $0$ & $\frac{6.40 }{ (15 - 6) } = 0.71$  \\

\end{tabular}
    }}%
    
\end{table*}

\begin{table*}[!ht]
    \centering
    \setlength{\tabcolsep}{0.5em}{\renewcommand{\arraystretch}{1.65}{%
    \footnotesize
    \caption{\label{tab:par_var_2pcoupled}
    Analogous to Table~\ref{tab:par_var_0pcoupled}, but for parametrization variations of coupled-channel $DD - DD^* - D^*D^*$  scattering in $J^P=2^+$ and $I=1$ (see Sec.~\ref{sec:2p}).
    The example parametrization (``$*$'')  is quoted in Eq.~\eqref{eq:2pcoupledpars}, and shown together with all other variations in Fig.~\ref{fig:coupled_2p}.
    In the cases where $\chisqadd$ was employed, the value of the parameter $E_l=0.64$ is used (see Eq.~\eqref{eq:chisqadd}).
    }
    
\begin{tabular}{c|c|c|c|c|c||c|c}
    \hline
    $D^*D^*$\SLJc{5}{S}{2}     & $DD^*$\SLJc{3}{D}{2}     & $DD$\SLJc{1}{D}{2}     & $DD$\SLJc{1}{D}{2}         & $DD^*$\SLJc{3}{D}{2}       & $DD^*$\SLJc{3}{D}{3}     & \multicolumn{2}{c}{\multirow{2}{4em}{$\chi^2/ N_\mathrm{dof}$}} \\
    $\to D^*D^*$\SLJc{5}{S}{2} & $\to DD^*$\SLJc{3}{D}{2} & $\to DD$\SLJc{1}{D}{2} & $\to D^*D^*$\SLJc{5}{S}{2} & $\to D^*D^*$\SLJc{5}{S}{2} & $\to DD^*$\SLJc{3}{D}{3} & \multicolumn{2}{c}{} \\
    
    \hline\hline
    \multicolumn{6}{c||}{Chew-Mandelstam phase space} & without $\chisqadd$ & with $\chisqadd$ \\\hline
    \bm{$0$} & \bm{$0$} & \bm{$0$} & \bm{$0$} & -- & \bm{$0$} & \bm{$\frac{15.53}{(20 - 5)} = 1.04 ^*$} & -- \\ 
    $0$ & $1$ & $1$ & $0$ & -- & $0$ & $\frac{18.33}{(20 - 5)} = 1.22$ & -- \\ 
    $0$ & $0$ & --  & $0$ & -- & $0$ & $\frac{16.60}{(20 - 4)} = 1.04$ & -- \\ 
    $0$ & $0$ & --  & $0$ & $0$ & $0$ & $\frac{11.66}{(20 - 5)} = 0.78$ & -- \\ 
    $0$ & $0$ & $0$  & -- & $0$ & $0$ & $\frac{10.80}{(20 - 5)} = 0.72$ & $\frac{11.00}{(20 - 5)} = 0.73$ \\ 
    $0$ & -- & $0$ & -- & $0$ & $0$ & $\frac{21.15}{(20 - 4)} = 1.32$ & -- \\ 
    $0$ & $0$ & $0$ & $0$ & $0$ & $0$ & $\frac{10.18}{(20 - 6)} = 0.73$ & -- \\ 
    $0$ & $0$ & $1$ & $1$ & $0$ & $0$ & $\frac{12.89}{(20 - 6)} = 0.92$ & -- \\ 

    \hline\hline
    \multicolumn{6}{c||}{Simple phase space} && \\\hline
    $0$ & $0$ & $0$ & $0$ & -- & $0$ & $\frac{16.37}{(20 - 5)} = 1.09$ & -- \\ 
    $0$ & $1$ & $1$ & $0$ & -- & $0$ & $\frac{19.28}{(20 - 5)} = 1.29$ & $\frac{19.57}{(20 - 5)} = 1.30$ \\ 
    $0$ & $0$ & -- & $0$ & -- & $0$ & $\frac{18.25}{(20 - 4)} = 1.14$ & $\frac{19.01}{(20 - 5)} = 1.19$ \\ 
    $0$ & $0$ & -- & $0$ & $0$ & $0$ & $\frac{12.61}{(20 - 5)} = 0.84$ & $\frac{16.37}{(20 - 5)} = 1.09$ \\ 
    $0$ & $0$ & $0$  & -- & $0$ & $0$ & $\frac{10.58}{(20 - 5)} = 0.71$ & $\frac{16.71}{(20 - 5)} = 1.11$ \\ 
    $0$ & -- & $0$ & -- & $0$ & $0$ & $\frac{23.34}{(20 - 4)} = 1.46$ & $\frac{24.41}{(20 - 4)} = 1.53$ \\ 
    $0$ & $0$ & $0$ & $0$ & $0$ & $0$ & $\frac{10.42}{(20 - 6)} = 0.74$ & $\frac{15.88}{(20 - 6)} = 1.13$ \\ 
    $0$ & $0$ & $1$ & $1$ & $0$ & $0$ & $\frac{12.78}{(20 - 6)} = 0.91$ & $\frac{16.59}{(20 - 6)} = 1.18$ \\ 

\end{tabular}

    }}%
\end{table*}

\clearpage

\twocolumngrid

\bibliography{refs.bib}
\bibliographystyle{apsrev}
\end{document}